\documentclass[a4paper,12pt]{article}

\usepackage{amsmath}
\usepackage{amssymb}
\usepackage[dvips]{graphicx}
\usepackage[dvips]{psfrag}

\newcommand{\be}{\begin{equation}}
\newcommand{\ee}{\end{equation}}
\newcommand{\bea}{\begin{eqnarray}}
\newcommand{\eea}{\end{eqnarray}}
\newcommand{\nono}{\nonumber}
\newcommand{\tr}{{\rm tr}}
\newcommand{\hg}{\hat{g}}
\newcommand{\vev}[1]{\left\langle #1 \right\rangle}

\newcommand{\Rnon}{R^{\rm non}}
\newcommand{\hR}{\hat{R}}
\newcommand{\Veff}{V_{\rm eff}}

\newcommand{\ha}{\hat{\alpha}}

\newcommand{\hP}{\hat{P}_*}
\newcommand{\hQ}{\hat{Q}_*}
\newcommand{\hT}{\hat{T}}
\newcommand{\huniv}[1]
{\left(#1+\sqrt{#1^2-\hT}\right)^{\frac{4}{3}}
+\left(#1-\sqrt{#1^2-\hT}\right)^{\frac{4}{3}}}
\newcommand{\cS}{{\cal S}}
\newcommand{\cO}{{\cal O}}
\newcommand{\cJ}{{\cal J}}
\newcommand{\prefI}{\frac{N}{h}}
\newcommand{\prefII}{\frac{N-1}{h'}}
\newcommand{\Rep}{{\rm Re}}
\newcommand{\Imp}{{\rm Im}}
\newcommand{\lam}{\lambda}
\newcommand{\der}{\partial}
\newcommand{\bare}{{\rm bare}}

\begin{document}
\begin{titlepage}
\begin{flushright}
\begin{tabular}{l}
KEK-TH-1118\\
OIQP-06-17\\
hep-th/0612042
\end{tabular}
\end{flushright}

\vspace{5mm}

\begin{center}
{\Large \bf T-duality of ZZ-brane}
\baselineskip=24pt

\vspace{15mm}
\large
Tsunehide Kuroki,\footnote{tkuroki@post.kek.jp}\\
{\it High Energy Accelerator Research Organization (KEK), \\
Tsukuba, Ibaraki 305-0801, Japan}\\
and\\
Fumihiko Sugino,\footnote{fumihiko\_sugino@pref.okayama.jp}\\
{\it Okayama Institute for Quantum Physics, \\
Kyoyama 1-9-1, Okayama 700-0015, Japan}
\vspace{20mm}
\end{center}
\begin{abstract}
We examine how nonperturbative effects in string theory 
are transformed under the T-duality in its nonperturbative framework 
by analyzing the $c=1/2$ noncritical string theory 
as a simplest example. We show that in the T-dual theory 
they also take the form of $\exp(-S_0/g_s)$ in the leading order 
and that the instanton actions $S_0$ of the dual ZZ-branes 
are exactly the same as those in the original $c=1/2$ string theory. Furthermore we present formulas for coefficients of 
$\exp(-S_0/g_s)$ in the dual theory.     
\end{abstract}
\end{titlepage}

\section{Introduction}
\label{sec:intro}

T-duality is a relation between perturbative vacua in string theory 
and is known to hold at each order in perturbation theory 
for the critical string theory \cite{Kikkawa:1984cp}. 
Since it is a characteristic feature of string theory, it is believed 
to play an important role even in constructing its nonperturbative 
formulation. From this point of view, it is quite intriguing 
to examine how nonperturbative effects are transformed under 
the T-duality in a possible nonperturbative framework of string theory. 
As such a framework, the noncritical string theory provides a useful 
toy model, because it can be formulated 
nonperturbatively by matrix models and their nonperturbative effects 
are identified as a kind of instanton effects 
\cite{David:1990sk}$\sim$\cite{Ishibashi:2005zf}. 
Moreover, we can explicitly formulate the T-duality 
in the noncritical string theory, for example, 
in the $c=1/2$ string theory as discussed in \cite{Asatani:1996jc}.  

However, as for the $c=1/2$ noncritical string theory, 
which does not have a continuous target space, 
it has been shown explicitly in \cite{Asatani:1996jc} 
that the theory is not invariant under 
the T-duality for lack of global winding modes 
associated to string world sheets of higher genus topology.  
This viewpoint for the T-duality still stays at the perturbative world sheet picture, 
although the corrresponding string field theory has been explicitly constructed. 
Therefore, as a first step to understand the T-duality truely at the nonperturbative level, 
it is important and interesting to identify 
nonperturbative effects in the dual $c=1/2$ string theory 
and to 
clarify how they correspond to those in the original 
$c=1/2$ string theory. This is the aim of this paper. 

Since as shown in refs.~\cite{Asatani:1996jc,Carroll:1995nj} 
the dual theory is defined by the $O(n)$ model 
on a random surface \cite{Duplantier:1988wc} with $n=1$,
we analyze this model and show that the leading nonperturbative effects 
in the dual theory also take the same form of $\exp (-S_0/g_s)$ 
as in the original theory, 
where $g_s$ is the string coupling constant. 
This form of the nonperturbative effects in general 
reflects the large-order behavior of perturbation series 
in string theory \cite{Shenker:1990uf}. 
Thus, we find that the perturbation series in the dual theory 
also shows the same large-order behavior. 
Moreover, for the standard $c<1$ noncritical string theory, 
the values of $S_0$ are deduced from the string equations 
\cite{David:1990sk,David:1992za,Eynard:1992sg}, or 
directly from matrix models as their instanton effects 
mentioned above. Other techniques deriving $S_0$ 
are also found in refs.~\cite{Fukuma:1996hj,deMelloKoch:2004en}. 
For the $O(n)$ model with general $n$ on a random surface, 
in contrast to the matrix models above,  
the string equation has not been derived (to the best of our knowledge), 
from which 
the nonperturbative effects in the dual theory can be seen. 
However, since the total free energies of 
the original theory and the dual theory are equivalent by definition 
for the case $n=1$\footnote
{Because of the subtlety of well-definedness for the matrix integrals 
in the double scaling limit, we can safely say that 
the equivalence holds in the sense of the $1/N$-expansion. 
}, 
the free energy of the dual theory should satisfy the same 
string equation as that of the original theory does. 
Therefore, we expect that the dual theory has exactly the same values 
of $S_0$ as those of the original theory. 
We will see that this is indeed the case. 
Since it is known in the standard noncritical string theories 
that $S_0$ can be identified 
\cite{McGreevy:2003kb}$\sim$\cite{Alexandrov:2003nn}
as the classical actions of the ZZ-branes \cite{Zamolodchikov:2001ah}, 
we can conclude that the classical actions of the dual ZZ-branes 
are the same as those of the original theory, and 
it gives an important basis in determining the T-duality transformation rule of the ZZ-branes. 
  
The organization of this paper is as follows. 
In the next section, 
we formulate the T-duality in the $c=1/2$ noncritical string theory 
in terms of the two-matrix model, and we see that the dual theory 
is defined by the $O(n)$ model on a random surface with $n=1$. 
In Section 3, we make the saddle point analysis in the $O(1)$ model 
and derive nonperturbative effects in the dual $c=1/2$ string 
theory. We proceed to examine the next-to-leading contribution 
to these nonperturbative effects in Section 4. The last section 
is devoted to the conclusions and discussions.     
In Appendix A, we explain the double scaling limits of the original and dual two-matrix models, 
and present a derivation of resolvents in the dual model. 
Some other computational details are discussed in Appendix B.

\section{T-duality in the $c=1/2$ noncritical string theory}
\label{sec:T-duality}
\setcounter{equation}{0}

The original $c=1/2$ string theory is defined 
by the double scaling limit \cite{Brezin:1990rb}
of a two-matrix model 
\begin{eqnarray}
Z & = & \int d^{N^2}Ad^{N^2}B \exp [-N\tr \, S(A,B)], \nonumber \\
S(A,B) & = & \frac{1}{2}A^2+\frac{1}{2}B^2
                       -cAB-\frac{g}{3}A^3-\frac{g}{3}B^3,
\label{original}                       
\end{eqnarray}
where $A$, $B$ are the $N\times N$ Hermitian matrices. 
This model describes the Ising model on a random surface 
\cite{Kazakov:1986hu} with the inverse temperature given by 
\begin{equation}
\beta=-\frac{1}{2}\log c.
\end{equation}
Various amplitudes in the $c=1/2$ string theory can be computed 
based on (\ref{original}), for example by solving the Schwinger-Dyson equations 
as discussed in refs.~\cite{Staudacher:1993xy,Sugino:1995hk}.
 
In order to perform the T-duality transformation to this model, 
we first note that 
the familiar T-duality transformation in closed string theories 
compactified on a circle is basically the same as 
the Kramers-Wannier dual transformation to the XY-model 
defined on the string world sheet (see, e.g. \cite{Shapere:1988zv} or Appendix A 
in \cite{Asatani:1996jc}). 
{}From this point of view, the T-duality transformation for this model 
will be naturally formulated as the Kramers-Wannier transformation 
of the Ising model on a random surface. 
Since the matrices $A$ and $B$ can be regarded as the up and down spins 
on a random surface, respectively, the Boltzmann factors 
of the original Ising model on a random surface are related 
to the bare propagators for $A$ and $B$ as
\begin{equation}
\vev{A_{ij}A_{kl}}_{\bare}=\vev{B_{ij}B_{kl}}_{\bare}=\frac{1}{N}\delta_{il}\delta_{jk} Le^{\beta},~~~
\vev{A_{ij}B_{kl}}_{\bare}=\frac{1}{N}\delta_{il}\delta_{jk}Le^{-\beta},
\end{equation}  
with $L=\sqrt{c}/(1-c^2)$. 
Therefore, the Boltzmann factors in the dual model should be 
proportional to $e^{\pm\tilde{\beta}}$, where $\tilde{\beta}$ 
is obtained by the $\mathbf{Z}_2$ Fourier transformation 
\begin{eqnarray}
e^{\beta}&=&K(e^{\tilde{\beta}}+e^{-\tilde{\beta}}), \nonumber \\
e^{-\beta}&=&K(e^{\tilde{\beta}}-e^{-\tilde{\beta}}).
\end{eqnarray}
$K$ is an overall normalization constant given by 
$K=(e^{2\tilde{\beta}}-e^{-2\tilde{\beta}})^{1/2}$. 
It is easy to introduce matrices realizing such bare propagators. 
If we define new matrix variables 
\begin{equation}
X=\frac{1}{\sqrt{2}}(A+B),~~~Y=\frac{1}{\sqrt{2}}(A-B),
\label{T-dual}
\end{equation}
then their bare propagators exactly give the Boltzmann factors 
of the dual model:
\begin{equation}
\vev{X_{ij}X_{kl}}_{\bare}=\frac{1}{N}\delta_{il}\delta_{jk}\frac{1}{\sqrt{1-c^2}}e^{\tilde{\beta}},~~~
\vev{Y_{ij}Y_{kl}}_{\bare}=\frac{1}{N}\delta_{il}\delta_{jk}\frac{1}{\sqrt{1-c^2}}e^{-\tilde{\beta}}.
\end{equation}
In this sense, these represent ``stick" or ``flip" 
of the dual spin, respectively. 
Thus, the T-duality transformation amounts to making a 
field redefinition (\ref{T-dual}) and 
the dual $c=1/2$ string theory is defined 
by the double scaling limit of a dual two-matrix model 
\cite{Asatani:1996jc,Carroll:1995nj}
\begin{eqnarray}
Z & = & \int d^{N^2}Xd^{N^2}Y \exp [-N\tr \, \tilde{S}_D(X,Y)], \nonumber \\
\tilde{S}_D(X,Y) & = & \frac{1-c}{2}X^2+\frac{1+c}{2}Y^2
                       -\frac{\hg}{3}\left(X^3+3XY^2\right), 
\label{dual}                       
\end{eqnarray}
where $\hg=g/\sqrt{2}$. 
This model is also known as the $n=1$ case of 
the $O(n)$ loop gas model on a random surface 
\cite{Duplantier:1988wc}.

The T-duality transformation (\ref{T-dual}) is a trivial change of the integration variables, 
and the total free energies given by 
\begin{equation}
F=-\frac{1}{N^2}\log Z,
\label{free energy}
\end{equation}
take the same value at least order-by-order in the $1/N$-expansions for both models. 
However, the coincidence of the free energies does not always 
lead to the T-duality. 
Indeed, we can see that the T-dual relation is broken in correlation functions on a surface of higher genus topology. 
For example, it is shown explicitly in \cite{Asatani:1996jc} 
that disk amplitudes in both theories have the same functional form, 
while a disk amplitude with one handle 
in the dual theory have different functional form 
from a corresponding amplitude in the original theory. 
More precisely, for the universal part of the disk and cylinder amplitudes, the following identification between the original and 
dual models holds:
\begin{eqnarray}
\frac{1}{\sqrt{2}}
\left(\tr\frac{1}{\zeta-A}+\tr\frac{1}{\zeta-B}\right)
\Longleftrightarrow
\tr\frac{1}{\xi-X},
\label{identification}
\end{eqnarray}
in the sense that they take the same functional form 
as functions of $\zeta$ and $\xi$ respectively 
with certain identifications of parameters. 
This is expected because both operators in (\ref{identification}) are interpreted as the 
Dirichlet type boundary conditions for the original 
and dual spins respectively, under the T-dual relation. 
Recall that $\vev{X_{ij}X_{kl}}_{\bare}$ 
represents ``stick" of the dual spin.  
On the other hand, the disk amplitude with one handle 
\begin{eqnarray}
\vev{\frac{1}{N}\tr\frac{1}{\xi-X}}_1
\end{eqnarray}
no longer has the same form as 
\begin{eqnarray}
\vev{\tr\left(\frac{1}{\zeta-A}+\frac{1}{\zeta-B}\right)}_1
\end{eqnarray} 
even for the universal part. 
(The subscript ``1'' put to the expectation values represents 
the random surface having one handle on which the expectation values are evaluated.) 
This difference originates from the excitations of odd number of $Y$-loops 
along a handle. Note that, if they are present, 
such a configuration cannot be interpreted as a dual spin configuration 
because $\vev{Y_{ij}Y_{kl}}_{\bare}$ represents the flip of the dual spin. 
Of course, from eqs.~(\ref{T-dual}) we have 
\be
\vev{\frac{1}{N}\tr\frac{1}{\xi-X}}_1=\vev{\frac{1}{N}\tr\frac{\sqrt{2}}{\sqrt{2}\xi-A-B}}_1.
\ee
However,  it does not imply the T-dual relation between the amplitudes, 
because the operator appearing in the r.h.s. is not consistent with the interpretation as ``stick" 
of the dual spin in the l.h.s. 
Thus, we should remark that, in order to show the T-duality, 
it is necessary to give a consistent interpretation of $X$ and $Y$ 
as dual spins, not only the transformation of variables (\ref{T-dual}). 
It is the same situation to the case of the Ising model on the regular lattice. 
The high temperature expansion of the Ising partition function on the plane has an one-to-one correspondence to 
the low temperature expansion. (For example, see \cite{Kogut:1979wt}.) 
It is a manifestation of the Kramers-Wannier duality (the T-duality). 
All the terms in the high temperature expansion can be written as configurations of loop gas. 
In the case of the surface with higher genus, some of them contain the loop gas configurations 
where odd number of the loops surround topologically nontrivial cycles.  
They can not be interpreted as dual spin configurations and violate the T-dual relation.  
Therefore, we can conclude that, although the total free energies are same between the original and dual theories, 
the T-duality does not hold in their higher genus parts due to the excitations of odd number of $Y$-loops along 
the handles.

In string theory with continuous target space, the T-dual symmetry arises when the target space is compactified.  
It is a symmetry under the interchange between momentum modes 
and winding modes in the compactified directions. 
{}From this viewpoint, the Ising model, whose target space is discrete (consists of two points), 
contains counterparts of the momentum modes, but not those of the winding modes 
for either case of random or regular lattice. 
We can understand that this asymmetry between the momentum and winding modes is 
the origin of the breaking of the T-duality\footnote{
It is possible to introduce the counterparts of the winding modes to matrix models with discrete target space 
to have the exact T-dual symmetry as discussed in \cite{KOSY}. 
}.  

In the original model given by (\ref{original}), 
we can apply the method of the orthogonal polynomial 
\cite{Itzykson:1979fi} and derive the string equation 
in the double scaling limit as 
\begin{equation}
f^3-\frac{3}{4}g_s^2ff^{\prime\prime}-\frac{3}{8}g_s^2(f')^2
+\frac{1}{24}g_s^4f^{(4)}=t,
\label{stringeq}
\end{equation}
where $t$ is the cosmological constant, $g_s$ is the string 
coupling constant, and $f(t)$ is the second derivative 
of the free energy $f(t)=g_s^2\ddot{F}(t)$ 
as a function of $t$ \cite{Brezin:1989db,Gross:1989ni}\footnote
{Note that we have changed the sign of the free energy 
compared to that in \cite{Brezin:1989db} and multiplied it 
by 2 because the potential in (\ref{original}) is not even.}. 
{}From this equation we can deduce the asymptotic expansion of $F$ 
as 
\begin{equation}
F(t)=\frac{9}{28}\frac{t^\frac{7}{3}}{g_s^2}+\frac{1}{24}\log t+\cdots, 
\label{genusexp}
\end{equation}
which is nothing but the genus expansion. 
Since as shown in Appendix \ref{app:dsl} 
(eqs.~(\ref{dsl_original}), (\ref{dsl_dual})) 
the double scaling limit is taken for the dual model~(\ref{dual}) in the same way 
as in the original model~\cite{Asatani:1996jc}, 
the free energy of the dual model should also satisfy the same 
string equation as in (\ref{stringeq}) in the double scaling limit\footnote
{In the dual model~(\ref{dual}), we have the expression of the integrals over the eigenvalues of 
$X$ and $Y$ (\ref{Z}) after integrating out 
the angle variables. It contains the factor $\prod_{i<j}1/(\mu_i+\mu_j)$, in addition to the 
Vandermonde determinants usually appearing in the case of the original model. 
Due to the additional factor, the orthogonal polynomial method does not work well, and 
it makes directly deriving the string equation difficult.
}   
and should have the same genus expansion as given in (\ref{genusexp}). 
However, even in this case, 
it is possible that they have different nonperturbative effects. 
Namely, suppose $f_1$ and $f_2$ satisfy the same string equation 
(\ref{stringeq}), 
then the semi-classical treatment of  (\ref{stringeq}) 
with $g_s \sim \hbar$ leads to 
the difference $\Delta f=f_1-f_2$ of the form:
\begin{equation}
\Delta f=C_1\frac{g_s^{\frac{1}{2}}}{t^{\frac{1}{4}}}
\exp\left(-\frac{6\sqrt{6}}{7g_s}t^{\frac{7}{6}}\right),~~~
\mbox{or}~~~ 
C_2\frac{g_s^{\frac{1}{2}}}{t^{\frac{1}{4}}}
\exp\left(-\frac{12\sqrt{3}}{7g_s}t^{\frac{7}{6}}\right),
\end{equation}
which yields a nonperturbative ambiguity of the free energy $F(t)$ in the following form:
\begin{equation}
\Delta F=\frac{C_1}{6}\frac{g_s^{\frac{1}{2}}}{t^{\frac{7}{12}}}
\exp\left(-\frac{6\sqrt{6}}{7g_s}t^{\frac{7}{6}}\right),~~~
\mbox{or}~~~ 
\frac{C_2}{12}\frac{g_s^{\frac{1}{2}}}{t^{\frac{7}{12}}}
\exp\left(-\frac{12\sqrt{3}}{7g_s}t^{\frac{7}{6}}\right).
\label{nonperturbative effect}
\end{equation}
$C_1$ and $C_2$ are numerical constants 
which cannot be determined from the string equation alone. 
Therefore, if we interpret one of $f_1$ and $f_2$ as a quantity of the original model and the other 
as that of the dual model, as long as $C_1$ or $C_2$ is nonzero, 
the dual theory has a different nonperturbative effect 
from that in the original theory. 
It is known that the nonperturbative effects 
in the original theory itself take the form 
(\ref{nonperturbative effect}) 
\cite{Kazakov:2004du,Ishibashi:2005zf}, 
so the dual theory must also have nonperturbative effects 
of the same form (up to the overall constants). 
We will see that this is indeed the case by computing 
the nonperturbative effects in the dual theory directly 
from the matrix model (\ref{dual}). 
It is well known that 
the exponents of the nonperturbative effects are identified as 
the actions of the ZZ-branes and are provided by the disk amplitudes 
in the presence of them.  Therefore, the fact that the dual theory 
also has the nonperturbative effects as 
in (\ref{nonperturbative effect}) implies that 
the actions of the ZZ-branes in the dual theory (the dual ZZ-branes) 
are the same as those in the original theory. 
It is worth noticing that the string equation 
can fix not only the exponents 
in the nonperturbative effects 
but the power of $t$ in the factors in front of them. 

On the other hand, the coefficients $C_1$ and $C_2$ in the nonperturbative effects cannot be fixed from the string equation. 
However, 
it is shown in \cite{Hanada:2004im,Ishibashi:2005zf} that, if we compute 
these coefficients in the $c<1$ noncritical string theory 
directly from a matrix model, they turn
 out 
to be unique and universal in the sense that they do not depend 
on details of a potential in the matrix model. Thus, it is interesting 
to examine whether this is also the case with 
the nonperturbative effects in the dual $c=1/2$ string theory 
and whether they agree with those in the original theory. 
Moreover, since the nonperturbative effects are 
computed from certain disk and cylinder amplitudes \cite{Hanada:2004im,Ishibashi:2005zf}, 
it will be possible to find out the T-duality at the nonperturbative level  
from the knowledge of the T-dual relation of the disk and cylinder amplitudes. 
It is expected that the analysis reveals the existence 
of large universality including T-duality 
for nonperturbative effects in string theory, 
or for string theory itself.


\section{Nonperturbative effects in the dual theory}
\label{sec:nonperturbative effect}
\setcounter{equation}{0}
In this section, we derive the leading part of the nonperturbative effects 
in the dual $c=1/2$ noncritical string theory 
directly by evaluating instanton contributions in the matrix model (\ref{dual}).

\subsection{Chemical potential of instanton}
\label{subsec:chemical potential}
As a preparation for instanton calculus in the dilute gas approximation,  
here we formulate the chemical potential of an instanton 
in the dual two-matrix model (\ref{dual}). 

By rescaling the matrices, the partition function in (\ref{dual}) 
becomes
\bea
Z_N(h)&=&\int dXdY \exp\left[-\prefI\tr \, S_D(X,Y)\right], 
\nonumber\\
S_D(X,Y)&=&\frac{1-c}{2}X^2+\frac{1+c}{2}Y^2
              -\frac{1}{3}\left(X^3+3XY^2\right), 
\label{dual2}
\eea
where $h=\hg^2$ and the measures $dX$, $dY$ are defined with the normalization:
\bea
\int dX 
\exp\left[-\frac{N}{h}
\tr\left(\frac{1-c}{2}X^2\right)\right]&=&1,\nonumber \\
\int dY 
\exp\left[-\frac{N}{h}
\tr\left(\frac{1+c}{2}Y^2\right)\right]&=&1,
\label{measure}
\eea 
so that the free energy 
has the standard $1/N$-expansion 
\be
F=-\frac{1}{N^2}\log Z_N(h) = F_0(h)+\frac{1}{N^2}F_1(h)+\frac{1}{N^4}F_2(h)+\cdots,
\label{Fexp}
\ee
where $F_k(h)$ represents the contribution from random surfaces with $k$-handles. 
   
The formula given in \cite{Itzykson:1979fi} enables us to 
rewrite the partition function in terms of the eigenvalues 
of $X$ and $Y$ as 
\be
Z_N(h) = D_N^{-1}\int \left(\prod_i d\lambda_id\mu_i \right)
 \frac{\Delta^{(N)}(\lambda)\Delta^{(N)}(\mu)}
      {\prod_{i<j}(\mu_i+\mu_j)} \exp\left[-\prefI \sum_i 
     \left(V(\lambda_i)+\frac{1+c}{2}\mu_i^2-\lambda_i\mu_i^2
     \right)\right], 
     \label{Z}
\ee
where $\lambda_i$, $\mu_i$ ($i=1,\cdots N$) are the eigenvalues 
of $X$ and $Y$ respectively, and 
$\Delta^{(N)}(\lambda)=\prod_{N\geq i>j\geq 1}(\lambda_i-\lambda_j)$ 
is the Vandermonde determinant. 
$V(\lambda)$ is defined as 
\be
V(\lambda)=\frac{1-c}{2}\lambda^2-\frac{1}{3}\lambda^3. 
\ee
The proportional constant $D_N$ is explicitly computed 
in Appendix \ref{app:denominator} (See eq. (\ref{D_Nexpre})). 

Hereafter, as a configuration of one instanton, 
we consider a situation where one pair of the eigenvalues 
(say ($\lambda_N$, $\mu_N$)) is separated from the other pairs ($\lam_i$, $\mu_i$) ($i=1, \cdots, N-1$) 
as a point on the $(\lambda,\mu)$-plane. Setting $x=\lambda_N$, $y=\mu_N$, 
the partition function is expressed as 
\bea
Z_N(h)
 & = & D_N^{-1}\int dxdy \int \left(\prod_{i=1}^{N-1}d\lam_id\mu_i \right)
       \frac{\Delta^{(N-1)}(\lam)\Delta^{(N-1)}(\mu)}
            {\prod_{i<j\le N-1}(\mu_i+\mu_j)}
       \prod_{i=1}^{N-1}\frac{(x-\lam_i)(y-\mu_i)}{(y+\mu_i)}
       \nono \\
 &   & \times e^{-\prefI \sum_{i=1}^{N-1} 
                  \left(V(\lam_i)+\frac{1+c}{2}\mu_i^2
                                    -\lam_i\mu_i^2\right)
          -\prefI \left(V(x)+\frac{1+c}{2}y^2-xy^2\right)}
     \nono \\
 & = & D_N^{-1}\int dxdy \, D_{N-1}\int dX'dY' \, 
       \frac{\det(x-X')\det(y-Y')}{\det(y+Y')} \, 
       e^{-\prefI \tr \, S_D(X',Y')}e^{-\prefI S_D(x,y)} \nono \\
 & = & \frac{D_{N-1}}{D_N}Z_{N-1}(h')
       \int dxdy 
       \vev{\frac{\det(x-X')\det(y-Y')}{\det(y+Y')}}' 
       e^{-\prefII S_D(x,y)} \nono \\
 & = & \frac{D_{N-1}}{D_N}Z_{N-1}(h')\int dxdy\,e^{-\Veff (x,y)},
 \label{isolation}
\eea
where $X'$, $Y'$ are $(N-1)\times (N-1)$ Hermitian matrices,  and 
\begin{eqnarray}
Z_{N-1}(h') & = & \int dX'dY' \, e^{-\prefII\tr \, S_D(X',Y')}, \nonumber \\
\vev{{\cal O}}' & = & 
\frac{1}{Z_{N-1}(h')} \int dX'dY' \, {\cal O} \, e^{-\prefII\tr \, S_D(X',Y')}.  
\end{eqnarray}
In the above equation, we have defined $h'$ as 
\be
\prefI=\prefII,
\label{h'}
\ee
and $\Veff(x,y)$ as 
\begin{equation}
e^{-\Veff(x,y)}\equiv
\vev{\frac{\det(x-X')\det(y-Y')}{\det(y+Y')}}' 
e^{-\prefII S_D(x,y)}. 
\label{Veff}
\end{equation}

As in the one-matrix model case considered in \cite{Hanada:2004im}, 
the total partition function is divided into multi-instanton sectors as 
\be
Z_N(h)=Z_N^{\rm (0-inst)}(h)+Z_N^{\rm (1-inst)}(h)+\cdots,
\ee
where the $k$-instanton sector is characterized by $k$-pairs of the eigenvalues 
separated from the other pairs on the $(\lam,\mu)$-plane. 
More precisely, 
the partition functions in the 1-instanton sector and 0-instanton 
sector are given by 
\bea
Z_N^{\rm (1-inst)}(h) & = & N\frac{D_{N-1}}{D_{N}}Z_{N-1}(h')
                   \int_{(x,y)\notin\cS} dxdy \,e^{-\Veff (x,y)},
\label{Z^1}                   
                   \\
Z_N^{\rm (0-inst)}(h) & = & \frac{D_{N-1}}{D_{N}}Z_{N-1}(h')
                   \int_{(x,y)\in\cS} dxdy \,e^{-\Veff (x,y)},
\label{Z^0}                                
\eea
respectively, where the factor $N$ in (\ref{Z^1}) 
means the number of ways to specify a separated pair of the eigenvalues, and 
$\cS$ is the support of the eigenvalue distribution on the 
$(\lambda,\mu)$-plane in the large-$N$ limit. Taking the ratio between them, 
we obtain 
\begin{equation}
\mu\equiv\frac{Z_N^{\rm (1-inst)}(h)}{Z_N^{\rm (0-inst)}(h)}
= N\frac{\int_{(x,y)\notin\cS} dxdy \,e^{-\Veff (x,y)}}
           {\int_{(x,y)\in\cS} dxdy \,e^{-\Veff (x,y)}}. 
\label{mu}           
\end{equation}
Following the argument given in \cite{Hanada:2004im}, 
it is easy to see that $\mu$ defined in this equation 
is in fact the chemical potential of the instanton, 
namely a statistical weight of the instanton, in the dilute gas approximation 
for the computation of the free energy 
of the dual two-matrix model. 

\subsection{Saddle point analysis}
\label{subsec:saddle}
In this subsection, we apply the saddle point method 
to the numerator in (\ref{mu}), which is valid in the large-$N$ limit. 
Since $\Veff(x,y)$ can be rewritten as 
\begin{eqnarray}
e^{-\Veff(x,y)} 
 & = & \vev{e^{\tr\log (x-X')+\tr\log (y-Y')-\tr\log (y+Y')}}'
          e^{-\prefII S_D(x,y)},
\end{eqnarray}
one may expect that 
$\Veff(x,y)$ in the large-$N$ limit can be expanded in terms of connected Green functions as 
\begin{align}
\lefteqn{e^{-\Veff (x,y)}} &&&\nonumber\\
 & = & \exp\Bigg[ &-\prefII S_D(x,y)
                 +\vev{\tr\log (x-X')}'_d+\vev{\tr\log (y-Y')}'_d
                 -\vev{\tr\log (y+Y')}'_d  
\nonumber \\
 &   & &+\frac{1}{2}\vev{(\tr\log (x-X'))^2}'_c
       +\frac{1}{2}\vev{(\tr\log (y-Y'))^2}'_c
       +\frac{1}{2}\vev{(\tr\log (y+Y'))^2}'_c \nonumber \\
 &   & &+\vev{\tr\log (x-X')\tr\log (y-Y')}'_c
        -\vev{\tr\log (y-Y')\tr\log (y+Y')}'_c \nonumber \\
 &   & &-\vev{\tr\log (x-X')\tr\log (y+Y')}'_c
        +{\cal O}\left(\frac{1}{N}\right)
             \Bigg],
\label{1/N-exp}             
\end{align}
where the subscripts ``$d$'' and ``$c$'' represent amplitudes of the disk and cylinder 
topologies, respectively. However, if $x$ or $y$ is inside 
the support of the eigenvalue distribution of $X'$ or $Y'$ 
respectively, the operator $\tr\log(x-X')$ or $\tr\log(y-Y')$ 
becomes large by itself to make the expansion (\ref{1/N-exp}) 
not valid \cite{Ishibashi:2005dh}. This motivates us 
to divide the numerator in (\ref{mu}) as 
\begin{eqnarray}
N\int_{(x,y)\notin\cS}dxdy \, e^{-\Veff (x,y)} & = &  
 N\int_{x\notin I_X, y\notin I_Y}dxdy \, e^{-\Veff (x,y)} \nono \\
 & & +N\int_{x\notin I_X, y\in I_Y}dxdy \, e^{-\Veff (x,y)}
+N\int_{x\in I_X, y\notin I_Y}dxdy \, e^{-\Veff (x,y)},\nonumber \\
\label{division}
\end{eqnarray}
where $\cS = I_{X} \times I_{Y}$, and $I_{X}$ ($I_{Y}$) is the support 
of the eigenvalue distribution of $X'$ ($Y'$).  
In our choice of the potential,  $I_{X}$ and $I_{Y}$ should be connected intervals, 
and their explicit forms are given in Appendix \ref{app:dsl}  as eqs. (\ref{cut_IX}) and (\ref{cut_IY}). 

\subsubsection{The first term in (\ref{division})}
First we consider the case where both $x$ and $y$ are outside the 
supports of the eigenvalue distributions of $X'$ and $Y'$. 
Then, the expansion~(\ref{1/N-exp}) is justified and 
the leading part of $\Veff(x,y)$ in the large-$N$ limit, denoted as $\Veff^{(0)}(x,y)$, is given by 
\begin{eqnarray}
\lefteqn{e^{-\Veff^{(0)}(x,y)}}\nonumber\\
 & = & 
\exp\Bigg[-\prefII S_D(x,y)
          +\vev{\tr\log (x-X')}'_d+\vev{\tr\log (y-Y')}'_d
          -\vev{\tr\log (y+Y')}'_d \Bigg]
 \nonumber \\
 & = & \exp\Bigg[-\prefII S_D(x,y)+\vev{\tr\log (x-X')}'_d \Bigg], 
 \label{e^Veff0}
\end{eqnarray}
where in the last line we have used the $\mathbf{Z}_2$ symmetry under 
$Y\rightarrow -Y$ of the action (\ref{dual2}). Therefore, we have 
\be
\Veff^{(0)}(x,y)=\prefII \left(V(x)+\frac{1+c}{2}y^2-xy^2\right)
                -\vev{\tr\log(x-X')}'_d. 
\label{Veff01}                
\ee
Since $\Veff^{(0)}(x,y)$ is proportional to $N-1$, we can apply 
the saddle point method to evaluate a leading contribution to the integrals 
of the first term in (\ref{division}) in the large-$N$ limit. 
The saddle point equations 
read 
\bea
0&=&\frac{\der \Veff^{(0)}(x,y)}{\der x}
 =\prefII\left(V'(x)-y^2-h'R_{X'}(x)\right),\nonumber\\
0&=&\frac{\der \Veff^{(0)}(x,y)}{\der y}
 =\prefII (1+c-2x)y, 
 \label{spe1}
\eea 
where $R_X(x)$ and $R_{X'}(x)$ are the resolvents for $X$ and $X'$, respectively: 
\be
R_{X}(x)=\vev{\frac{1}{N}\tr\frac{1}{x-X}}_d, \qquad 
R_{X'}(x)=\vev{\frac{1}{N-1}\tr\frac{1}{x-X'}}'_d.
\label{resolvent}
\ee  
In the large-$N$ limit, these two become coincident, 
and the explicit form is given by eqs. (\ref{RXnon_final}) and (\ref{RXuniv_final}) in Appendix \ref{app:dsl}. 
Since it is seen from (\ref{cut_IY}) that the origin $y=0$ belongs to $I_Y$ 
(which is also suggested by the $\mathbf{Z}_2$ symmetry), 
the second equation in (\ref{spe1}) leads to the solution 
$x_0=(1+c)/2\equiv\hP$ 
as a saddle point of $x$. 
This value coincides with the critical point of $x$, 
at which a cubic equation satisfied by the universal part of the resolvent $\hat{R}_X(x)$ 
becomes triply degenerate as 
$\hat{R}_X(x_0)^3=0$ when $\hat{g} = \hat{g}_*$, $c=c_*$. 
(For details, see ref.~\cite{Asatani:1996jc} or Appendix~\ref{app:dsl})\footnote
{Hereafter, in taking the large-$N$ limit, $c$ is fixed at the critical value $c_* = \frac{-1+2\sqrt{7}}{27}$.
}.
The nonuniversal part of $R_X(x)$ is given by 
\be
\Rnon_X(x)=\frac{1}{3h}(2V'(x)-V'(1+c-x)),~~~
R_X(x)=\Rnon_X(x)+\hR_X(x). 
\label{nonuniversal}
\ee
Making use of (\ref{nonuniversal}) to the first equation in (\ref{spe1}) determines
saddle points of $y$ as 
\be
y_0=\pm 2c\equiv\pm\hQ,
\ee
which respects the $\mathbf{Z}_2$ symmetry. 
{}From eqs. (\ref{cut_IX}), (\ref{cut_IY}) 
in Appendix \ref{app:dsl}, we recognize these saddle points to be 
outside the supports of the eigenvalue distributions 
in the double scaling limit:\footnote
{It turns out that the $o(a)$ term of $\gamma$ in (\ref{cut_IY}) is positive, 
by considering next-to-leading contributions to the solution of (\ref{SD_RY3}). 
Thus, strictly speaking, $y_0$ is slightly inside the cut $I_Y$ by the order of $o(a)$, 
and coincides with the right or left edge of $I_Y$ 
in the double scaling limit. Hence there is some subtlety 
in justification of $\Veff^{(0)}(x,y)$ given in (\ref{Veff01}). 
However, since the quantity $o(a)$ is in fact negligible compared to contributions of $\cO(a)$ usually playing 
relevant roles in the double scaling limit, 
we may assume that the eigenvalue distribution at $y_0$ is almost zero, 
and that the operator $\tr\log (y_0-Y)$ is not so singular 
that it can invalidate (\ref{Veff01}).}
\be
x_0=\hP\notin I_X,~~~y_0=\pm\hQ\notin I_Y.
\label{sp1}
\ee
At the saddle points, $\Veff^{(0)}(x,y)$ in (\ref{Veff01}) takes the form 
\be
\Veff(\hP,\pm\hQ)=\prefII V(\hP)-\vev{\tr\log(\hP-X')}'_d.
\label{spvalue1}
\ee

\subsubsection{The second term in (\ref{division})}
For the case $y\in I_Y$, we have the solution $y_0=0\in I_Y$ in the 
second equation of  (\ref{spe1}). 
The solution is also expected from the $\mathbf{Z}_2$ symmetry.  
However, as noticed above, in this case 
we cannot trust the form of $\Veff^{(0)}(x,y)$ given in (\ref{Veff01}). 
This kind of saddle point first appears in the dual model, 
not seen in the case of the original model~\cite{Kazakov:2004du,Ishibashi:2005zf}. 
In order to resolve this problem, we consider 
performing the $Y$-integration first in the partition function 
and then calculating instanton effects from isolated eigenvalues in the resulting 
one-matrix model for $X$. In evaluating the second term 
of (\ref{division}), we replace the integration over the interval $I_Y$ 
with that over the whole real axis. Then, saddle points 
in the $y\notin I_Y$ region, which are nothing but the ones 
considered in (\ref{sp1}), would give rise to an error in this 
replacement. However, as shown in Section~\ref{sec:case_1}, 
they only give 
contributions exponentially small by the factor (\ref{ourNPE1}) 
compared to those from the integration over $y\in I_Y$. 
Thus, concerning the leading 
contribution of the second term in (\ref{division}), we can neglect such an error and 
justify the replacement of the integration region. 

We go back to the expression (\ref{isolation}) 
to rewrite the second term in (\ref{division}) as  
\bea
\lefteqn{N\int_{x\notin I_X, y\in I_Y} dxdy \, e^{-\Veff(x,y)}}\nono \\ 
&=&\frac{D_{N-1}^{-1}}{Z_{N-1}(h')}\int_{x\notin I_X}dx
   \int \left(\prod_{i=1}^Nd\lam_id\mu_i\right) \sum_{i=1}^N\delta(x-\lam_i) \, 
   \frac{\Delta^{(N)}(\lam)\Delta^{(N)}(\mu)}{\prod_{i<j\le N}(\mu_i+\mu_j)}
        \, e^{-\prefII\sum_{i=1}^{N}S_D(\lam_i,\mu_i)}\nono \\
&=&\frac{D_{N-1}^{-1}}{Z_{N-1}(h')}\int_{x\notin I_X}dx \, D_N
   \int dXdY \, \tr\, \delta(x-X) \, e^{-\prefI\tr \, S_D(X,Y)}, 
\eea   
where we have inserted $1=\int d\lambda_N \, \delta (x-\lambda_N)$ 
and used the fact that the integrand is symmetric 
under the interchange among $\{\lam_i,x\}$ ($i=1, \cdots, N-1$). 
The measures $dX$, $dY$ are normalized as (\ref{measure}). 
As the result of the $Y$-integration, we obtain 
\bea
\lefteqn{N\int_{x\notin I_X, y\in I_Y} dxdy\, e^{-\Veff(x,y)}}\nonumber\\
&=&\frac{D_{N-1}^{-1}D_N}{Z_{N-1}(h')}
   \int_{x\notin I_X}dx
   \int dX \, \tr\, \delta(x-X)\nono \\
& &\times\exp\left[-\prefI\tr V(X)
             -\frac{1}{2}\tr\log
              \left(\mathbf{1}\otimes\mathbf{1}
             -\frac{1}{1+c}(X\otimes\mathbf{1}
                           +\mathbf{1}\otimes X)\right) \right]. 
                           \label{Yint}
\eea
Next, integration over the angular variables of $X$ yields 
\bea
\lefteqn{N\int_{x\notin I_X, y\in I_Y} dxdy \, e^{-\Veff(x,y)}}\nonumber\\
&=&
\frac{D_{N-1}^{-1}D_N}{Z_{N-1}(h')}
   \int_{x\notin I_X}dx \, 
   {J_N^X}^{-1}\int \left(\prod_{i=1}^Nd\lam_i\right) \, \Delta^{(N)}(\lam)^2
   \sum_{i=1}^N\delta(x-\lam_i)\nono \\
& &\times\exp\left[-\prefI\sum_{i=1}^N V(\lam_i)
            +\frac{1}{2}\sum_{n=1}^{\infty}\frac{1}{n}
     \left(\frac{1}{1+c}\right)^n\sum_{k=0}^n{}_nC_k
     \sum_{i=1}^N \lam_i^k\sum_{j=1}^N \lam_j^{n-k}\right],
\label{angularint}          
\eea
where the normalization constant $J_N^X$ arises upon the angular integration and 
satisfies for an arbitrary $U(N)$-invariant function of $X$: $f(\tr X, \tr X^2, \cdots)$ 
\be
J_N^X\int dX \, f(\tr X, \tr X^2, \cdots)
=\int \left(\prod_{i=1}^Nd\lambda_i\right) \, \Delta^{(N)}(\lambda)^2\, 
f\left(\sum_i\lambda_i, \sum_i\lambda_i^2, \cdots\right). 
\label{J^X}
\ee
$J_N^X$ is calculated in Appendix \ref{app:denominator} as
(\ref{J_N^X}). 
We rename one of  $\lambda_i$'s as $x$ by integrating the delta functions $\sum_{i=1}^N\delta(x-\lam_i)$ 
and introduce the $(N-1)\times (N-1)$ Hermitian matrix $X'$ again 
to obtain the effective potential $\Veff(x)$: 
\bea
\lefteqn{N\int_{x\notin I_X, y\in I_Y} dxdy \, e^{-\Veff(x,y)}}\nonumber\\
&=&N\frac{D_{N-1}^{-1}D_N{J_N^X}^{-1}}{Z_{N-1}(h')}
   \int_{x\notin I_X}dx \, J_{N-1}^X
   \int dX' \det (x-X')^2\nono \\
 &&\times\exp\Biggl[-\prefII \tr V(X')-\prefII V(x)
  -\frac{1}{2}\log\left(1-\frac{2x}{1+c}\right)
  -\tr\log\left(1-\frac{x+X'}{1+c}\right)\nono \\
 &&\hspace{1.2cm}
  -\frac{1}{2}\tr\log
   \left(\mathbf{1}\otimes\mathbf{1}
             -\frac{1}{1+c}(X'\otimes\mathbf{1}
                           +\mathbf{1}\otimes X')\right)\Biggr]\nono\\
&=&N\frac{D_{N-1}^{-1}D_N{J_N^X}^{-1}J_{N-1}^X}{Z_{N-1}(h')}
  \int_{x\notin I_X}dx
   \int dX'dY' \det (x-X')^2 \, (c+1)^{N-\frac{1}{2}}\nono \\
 &&\times\exp\Biggl[-\prefII V(x)
  -\frac{1}{2}\log(c+1-2x)
  -\tr\log(c+1-x-X')\Biggr]\nono \\
 &&\times\exp\left[-\prefII\tr
        \left(V(X')+\frac{1+c}{2}Y^{\prime 2}-X'Y^{\prime 2}
        \right)\right]\nono\\ 
&=&N\frac{D_NJ_{N-1}^X}{D_{N-1}J_N^X}
   (c+1)^{N-\frac{1}{2}}
   \int_{x\notin I_X}dx
   \vev{\frac{\det (x-X')^2}{\det (c+1-x-X')}}'
 e^{-\prefII V(x)-\frac{1}{2}\log(c+1-2x)}
\nono\\
&\equiv&N\frac{D_NJ_{N-1}^X}{D_{N-1}J_N^X}
   (c+1)^{N-\frac{1}{2}}
   \int_{x\notin I_X}dx \, e^{-\Veff(x)}.
\label{O(n)num2}   
\eea
Therefore, when $y\in I_Y$, the effective potential 
for $x\notin I_X$ is expressed as 
\be
e^{-\Veff(x)}=\vev{\frac{\det (x-X')^2}{\det (c+1-x-X')}}'
 e^{-\prefII V(x)-\frac{1}{2}\log(c+1-2x)}. 
\ee
In the large-$N$ limit, the leading term of $\Veff(x)$ is reduced to 
\be
\Veff^{(0)}(x)=\prefII V(x)-2\Rep\vev{\tr\log (x-X')}'_d
                     +\vev{\tr\log(c+1-x-X')}'_d.
\label{Veff02}                     
\ee
Note that for large $x\notin I_X$, 
the third term may yield the imaginary part which prevents us 
from interpreting $\Veff^{(0)}(x)$ as the effective potential. 
However, at least near the saddle point 
for $x$ to be determined in Section~\ref{sec:case_2}, 
the point $c+1-x$ is also outside the cut and the third term 
stays real. Since $\Veff^{(0)}(x)$ is again proportional to $N-1$, 
we can apply the saddle point method to compute 
the second term of (\ref{division}) in the large-$N$ limit. 
 
The saddle point equation from $\Veff^{(0)}(x)$ reads 
\be
0=\frac{\partial\Veff^{(0)}(x)}{\partial x}
=-(N-1)\left(2\Rep \, \hR_{X'}(x)+\hR_{X'}(c+1-x)\right),
\label{spe2}
\ee
where 
we have used (\ref{nonuniversal}). 
The saddle point equation (\ref{spe2}) is solved 
in the double scaling limit in the next subsection. 

Finally we comment on the third term in (\ref{division}). 
The solutions (\ref{sp1}) for the equations (\ref{spe1}) are unique in the case $y\notin I_Y$, 
and thus (\ref{Veff01}) has no saddle point in the region $x\in I_X$, $y\notin I_Y$. 
Although strictly speaking 
the expression of the effective potential in the large-$N$ limit 
given in (\ref{Veff01}) can not be sufficiently trusted when $x\in I_X$, $y\notin I_Y$, 
here we 
assume that the third term in (\ref{division}) does not contribute 
to the numerator in (\ref{mu}). 

\subsection{Nonperturbative effects}
The leading term of the denominator in (\ref{mu}) is computed as 
\bea
\lefteqn{\left.\int_{(x,y)\in\cS} dxdy \,e^{-\Veff (x,y)}
\right|_{\text{leading}}} \nono \\
&  = &  \exp\left[\vev{\tr\log(\beta-X')}'_d
      +(N-1)\int_{c+1-\beta}^{\beta}R_{X'}(x')dx'
      -\prefII V(\beta)\right], 
\label{den}      
\eea
where $\beta$ is the right edge of the eigenvalue distribution 
of $X$ in the one-matrix model appearing in (\ref{Yint}): 
\be
Z_N(h)=\int dX \exp\left[-\prefI\tr \, V(X)
             -\frac{1}{2}\tr\log
              \left(\mathbf{1}\otimes\mathbf{1}
             -\frac{1}{1+c}(X\otimes\mathbf{1}
                           +\mathbf{1}\otimes X)\right) \right]. 
\label{Z_O(1)}                           
\ee 
Since this model is obtained simply by performing the Gaussian integration 
over $Y$ in the dual model (\ref{dual2}), the partition 
function is exactly the same as in (\ref{dual2}). 
Details in the derivation of (\ref{den}) are given in 
Appendix \ref{app:denominator} (See eq.~(\ref{denfinal})). 

\subsubsection{The case $x_0\not\in I_X$, $y_0\not\in I_Y$}
\label{sec:case_1}
At the saddle points (\ref{sp1}), the leading term of the numerator 
in (\ref{mu}) is given by the effective potential (\ref{spvalue1}). 
Putting it together with the contribution from the denominator (\ref{den}), 
the leading term of the chemical potential $\mu$ takes the form as
\begin{align}
(\mu~{\rm leading})
=\exp\Biggl[&-\prefII \left(V(\hP)-V(\beta)\right)\nono \\         
       &+(N-1)\left(\int_{\beta}^{\hP}R_{X'}(x')dx'
        -\int_{c+1-\beta}^{\beta}R_{X'}(x')dx'\right)\Biggr].
\end{align}
Note that the potential terms above exactly cancel the nonuniversal parts of the resolvents, 
by using the expression (\ref{nonuniversal}), to give 
\be
(\mu~{\rm leading})
=\exp\Biggl[(N-1)\left(\int_{\beta}^{\hP}\hR_{X'}(x')dx'
        -\int_{c+1-\beta}^{\beta}\hR_{X'}(x')dx'\right)\Biggr].
        \label{muleading1}
\ee
It can be written solely in terms of the universal part of the resolvent, 
and the nonuniversal part does not contribute. 
This observation would be relevant 
if we consider proving the universality of nonperturbative effects 
in the dual theory as done for the standard 
$c<1$ noncritical string theories 
\cite{Hanada:2004im,Kawai:2004pj,Ishibashi:2005zf}. 

In order to take the double scaling limit, let us introduce the scaling variable $\zeta$ 
as $x'=\hP(1+a\zeta)$ with the lattice constant $a$. Then, the resolvent is expressed as 
\be
\hR_X(x')=a^\frac{4}{3}\ha w(\zeta) + \cO\left(a^\frac{5}{3}\right),~~~
w(\zeta)=\huniv{\zeta}, 
\label{w}
\ee
where $\ha$ is a numerical constant and 
$\hT$ is proportional to the cosmological constant $t$ 
defined by 
\be
\ha = \frac{\hat{s}^\frac{4}{3}}{2^\frac{2}{3}\cdot 5c}, \qquad \hg=\hg_*\left(1-a^2t\right)
   =\hg_* \left(1-a^2\frac{\hat{s}^2}{10}\hT\right). 
\ee
$\hat{s}$ is an irrational number $\hat{s} = 1+\sqrt{7}$, and $\hg_*=\sqrt{5c^3}$ is the critical point of $\hg$. 
(For more details in the double scaling limit of the dual model, 
see Appendix \ref{app:dsl}.) 
We carry out the integrations in (\ref{muleading1}) by moving to the variable $\zeta$ as
\bea
x'=\hP &\Leftrightarrow& \zeta=0,\nono \\
x'=\beta=\hP\left(1-a\sqrt{\hT}\right) &\Leftrightarrow& \zeta=-\sqrt{\hT}, 
\nono \\
x'=c+1-\beta=2\hP-\beta &\Leftrightarrow& \zeta=\sqrt{\hT},
\eea
and end up with the result
\be
(\mu~{\rm leading})
=\exp\left(-5^{\frac{1}{6}}
 \frac{12\sqrt{6}}{7}(N-1)a^{\frac{7}{3}}t^{\frac{7}{6}}\right). 
 \label{ourNPE1}
\ee
Since the sphere free energy is expressed as
\be
N^2 F_0=\frac{9}{7} \, 5^\frac{1}{3}N^2a^\frac{14}{3}t^\frac{7}{3} 
\label{sphere_F}
\ee
in the parametrization used here\footnote
{Eq. (\ref{sphere_F}) is obtained, for example, 
by integrating the leading term of  $\hat{W}_3$ 
in (\ref{W3}) with respect to $t$, since 
\be
W_3 = \vev{\frac{1}{N}\tr\, A^3}_d = -\frac32\frac{\partial F_0}{\partial g} 
\ee
in the original model, and the spherical free energy takes the same form for both of the original and dual models.
}, 
we take the double scaling limit in such a way that 
(\ref{sphere_F}) agrees with the first term 
in (\ref{genusexp}). 
Explicitly, 
\be
N \to \infty \quad {\rm and} \quad a\to 0 \quad {\rm with} \quad 
\frac{1}{g_s}\equiv 2\cdot 5^{\frac{1}{6}}Na^{\frac{7}{3}} \quad {\rm fixed}. 
\label{dsl_gs}
\ee
Then, $(\ref{ourNPE1})$ becomes 
\be
(\mu~{\rm leading})
=\exp\left(-\frac{6\sqrt{6}}{7g_s}t^{\frac{7}{6}}\right),
\label{leadingNPE1}
\ee
which exactly reproduces one of the leading terms 
in the nonperturbative effects given in (\ref{nonperturbative effect}). 
Moreover, we have found two degenerate saddle points (\ref{sp1}) 
which give this nonperturbative effect. 
Notice that 
the original $c=1/2$ theory also has two degenerate nonperturbative effects 
exactly given by (\ref{leadingNPE1})
\cite{Kazakov:2004du,Ishibashi:2005zf}. 
Thus, the leading nonperturbative effects (\ref{leadingNPE1}) 
precisely match between the original theory and the dual one 
including their multiplicity.

\subsubsection{The case $x_0\not\in I_X$, $y_0\in I_Y$}
\label{sec:case_2}
{}From eqs.~(\ref{Veff02}) and (\ref{den}), 
the leading contribution to $\mu$, which comes from a saddle point $x=x_0\notin I_X$ satisfying (\ref{spe2}),
takes the form 
\begin{align}
(\mu~{\rm leading}) &=&
\exp\Biggl[
&-\prefII \left(V(x_0)-V(\beta)\right)\nono \\
&&&+(N-1)\left(2\int_{\beta}^{x_0}R_{X'}(x')dx'
            +\int_{\beta}^{x_0}R_{X'}(c+1-x')dx'\right)
            \Biggr].
\end{align}
It is again expressed only by the universal part:
\be
(\mu~{\rm leading}) 
=\exp\left[(N-1)\int_{\beta}^{x_0}(2\hR_{X'}(x')+\hR_{X'}(c+1-x'))dx'\right],
\label{mu2final}
\ee 
as anticipated from (\ref{spe2}).  
Note that both of $R_{X'}(x')$ and $R_{X'}(c+1-x')$ do not have the imaginary 
parts for $\beta\leq x'\leq x_0$. In terms of the scaling variable $\zeta$, 
the saddle point equation (\ref{spe2}) is written as 
\be
2w(\zeta)+w(-\zeta)=0,
\ee
and it is easy to see that a solution to this equation 
on the first (physical) sheet is given by $\zeta_0=\sqrt{\hT/2}$. 
Substituting this value into (\ref{mu2final}),  
we find another nonperturbative effect 
\be
(\mu~{\rm leading}) =\exp\left(
-5^{\frac{1}{6}}\frac{24\sqrt{3}}{7}(N-1)t^{\frac{7}{6}}a^\frac{7}{3}
\right).
\ee
In the double scaling limit (\ref{dsl_gs}), this becomes 
\be
(\mu~{\rm leading}) =\exp\left(
-\frac{12\sqrt{3}}{7g_s}t^{\frac{7}{6}}\right), 
\label{leadingNPE2}
\ee
to reproduce the leading term of the remaining 
nonperturbative effect in (\ref{nonperturbative effect}). 
It is also exactly the same as the remaining 
one of three nonperturbative effects in the original 
$c=1/2$ string theory. 
Combining the results in (\ref{leadingNPE1}) and (\ref{leadingNPE2}), 
we conclude that the T-duality transformation 
does not change the number of the ZZ-branes 
and their actions (weights). 

Before closing this section, we make the following one comment. 
The disk amplitude $\vev{\tr \log (y-Y')}_d'$ (or $R_{Y'}(y)$) represents a quite 
singular dual spin configuration along the boundary, and its counterpart 
in the original model does not appear in the instanton calculus~\cite{Ishibashi:2005zf}. 
It is worthy of noticing that such amplitudes cancel out due to the ${\mathbf Z}_2$ symmetry 
in the process of the computation of $\Veff^{(0)}(x,y)$ 
in (\ref{e^Veff0}) and they do not affect the interpretation of the dual ZZ-branes.

\section{Coefficients of the nonperturbative effects}
\label{sec:coefficient}
\setcounter{equation}{0}
In this section, we consider next-to-leading terms 
in the chemical potential of the instanton, namely, 
coefficients in front of the nonperturbative effects 
(\ref{leadingNPE1}) and (\ref{leadingNPE2}). 

The coefficient coming from the denominator in (\ref{mu}) 
is evaluated in Appendix \ref{app:denominator}.  
In eq.~(\ref{denominator_final}), the first term in the exponential 
$2(N-1)R$ represents nothing but the leading term (\ref{den}), 
and the remaining parts 
\be
(2\pi)^{\frac32} \sqrt{(N-1)h'} \, \exp\left(R+h'\frac{\der R}{\der h'}\right)
\ee
contribute to the coefficient. 
$R$ is defined by eq.~(\ref{R_def}), and $R+h'\frac{\der R}{\der h'}$ can be written in terms of 
several disk and cylinder amplitudes as (\ref{denominator coeff}). 

In the following, let us consider contributions to the coefficient from the numerator 
in each of the cases $x_0\not\in I_X$, $y_0\not\in I_Y$ and $x_0\not\in I_X$, $y_0\in I_Y$. 

In the first case $x_0\not\in I_X$, $y_0\not\in I_Y$, 
the next-to-leading contributions are provided by the ${\cal O}(N^0)$ 
part in the exponent in (\ref{1/N-exp}) evaluated at the saddle point 
and also by the Gaussian integration of (\ref{e^Veff0}) 
around the saddle point. 
The latter is easy to calculate to yield 
for both of $(x_0, y_0)=(\hP, \pm\hQ)=(\frac{1+c}{2}, \pm 2c)$ 
\be
N\int_{x\not\in I_X, y\not\in I_Y} dxdy \, e^{-\Veff(x,y)}
=i\frac{\pi h}{2c} \, e^{-\Veff^{(1)}(x_0,y_0)}e^{-\Veff^{(0)}(x_0,y_0)}
\left(1+{\cal O}\left(\frac{1}{N}\right)\right),
\label{gauss1}
\ee
where 
\bea
\Veff^{(0)}(x, y) & = &\prefII \left(V(x)+\frac{1+c}{2}y^2-xy^2\right)
                -\vev{\tr\log(x-X')}'_d,\nonumber\\
\Veff^{(1)}(x,y) & = &
       -\frac{1}{2}\vev{(\tr\log (x-X'))^2}'_c
       -\vev{(\tr\log (y-Y'))^2}'_c \nono \\
 & &  +\vev{\tr\log (y-Y')\tr\log (y+Y')}'_c
 \label{trlog1}
\eea
are obtained from (\ref{1/N-exp}) 
by using the $\mathbf{Z}_2$ symmetry. 
For both of the two saddle points (\ref{gauss1}) takes the same value, 
which should be expected from the $\mathbf{Z}_2$ symmetry. 
Together with the contribution from the denominator, 
$e^{-\Veff^{(0)}(x_0,y_0)-2(N-1)R}$ gives the leading part (\ref{leadingNPE1}), and  
$\Veff^{(1)}(x,y)$ represents the next-to-leading term in $\Veff (x,y)$. 
The cylinder amplitudes appearing in $\Veff^{(1)}(x,y)$ 
will be found by integrating with respect to $z$ and $z'$ 
\be
\vev{\tr\frac{1}{z-X'}\tr\frac{1}{z'-X'}}_c,~~~
\vev{\tr\frac{1}{z-Y'}\tr\frac{1}{z'-Y'}}_c,~~~
\vev{\tr\frac{1}{z-Y'}\tr\frac{1}{-z'-Y'}}_c.
\label{cylinder amplitudes}
\ee
For this purpose, 
we will need the explicit form of these cylinder amplitudes 
without taking the double scaling limit. 
In fact, as long as the first one in (\ref{cylinder amplitudes}) 
is concerned, such an amplitude is calculated in \cite{Eynard:1995nv}. 
However, the result has a complicated form to 
makes it difficult to perform the $z,z'$-integration explicitly. 
Moreover, to the best of our knowledge, the cylinder amplitudes 
for $Y'$ like the remaining two in (\ref{cylinder amplitudes}) 
have not been found in the literature. 

Similarly, in the second case $x_0\not\in I_X$, $y_0\in I_Y$, 
from (\ref{O(n)num2}) the numerator takes the form of 
\bea
\lefteqn{N\int_{x\not\in I_X, y\in I_Y}
dx \, e^{-\Veff (x)}}\nonumber\\
&=&N\frac{D_NJ^X_{N-1}}{D_{N-1}J_N^X}(c+1)^{N-\frac{1}{2}}
\frac{1}{\sqrt{c+1-2x_0}}
\sqrt{\frac{2\pi}{\Veff^{(0)\prime\prime}(x_0)}} \, 
e^{-\Veff^{(1)}(x_0)}e^{-\Veff^{(0)}(x_0)}
\left(1+{\cal O}\left(\frac{1}{N}\right)\right),\nonumber\\
\label{spmethod2}
\eea
where $x_0 = \hP \left(1+a\zeta_0\right)=\hP \left(1+a\sqrt{\hT/2}\right)$, 
\bea
\Veff^{(0)}(x)&=&\prefII V(x)-2\Rep\vev{\tr\log (x-X')}'_d
                     +\vev{\tr\log(c+1-x-X')}'_d, \nonumber\\
\Veff^{(1)}(x)&=&-\frac{1}{2}\vev{(\tr\log (x-X')^2)^2}'_c
-\frac{1}{2}\vev{(\tr\log (c+1-x-X'))^2}'_c \nonumber \\
&&+\vev{\tr\log (x-X')^2\, \tr\log (c+1-x-X')}'_c, 
\label{trlog2}
\eea
and $e^{-\Veff^{(0)}(x_0)-2(N-1)R}$ provides the leading contribution (\ref{leadingNPE2}). 
The factor coming from the Gaussian integration can be 
evaluated easily in the double scaling limit. Then we find 
\be
N\int_{x\not\in I_X, y\in I_Y}
dxdy \, e^{-\Veff (x,y)} =\sqrt{\frac{\sqrt{3}\pi^2h}{4\hat{\alpha}
      a^\frac{4}{3}\hT^\frac{2}{3}}} \, 
e^{-\Veff^{(1)}(x_0)}e^{-\Veff^{(0)}(x_0)}.
\ee
Note that the factor has the $a$-dependence of $a^{-\frac23}$ and it is real.  
Therefore, it is important to check that $e^{-\Veff^{(1)}(x_0)}$ 
exactly cancels the $a$-dependence in the prefactor 
so that the nonperturbative effect will be finite 
in the double scaling limit. 
Also, it is interesting to see whether $e^{-\Veff^{(1)}(x_0)}$ provides 
the factor $i$. 
In examining $\Veff^{(1)}(x_0)$, however we encounter again 
the technical problem in the $z, z'$-integration of 
the first amplitude 
in (\ref{cylinder amplitudes}).

\section{Conclusions and discussions}
\label{sec:discussions}
\setcounter{equation}{0}

Based on the dual two-matrix model,  
we formulated the chemical potential of the instanton and 
showed that the instantons in the dual model give the same 
nonperturbative effects as those in the original model 
in the leading order. 
Since the dual model defines the T-dual theory 
of the standard $c=1/2$ noncritical string theory, 
it implies that we have identified 
the degrees of freedom of the dual ZZ-branes, 
which provide nonperturbative 
effects in the dual theory, and shown that 
the number of species of the dual ZZ-branes and their tensions 
coincide with the original ones. Thus our result gives 
further support on the identification of the ZZ-brane 
or nonperturbative effect in string theory 
as the instanton in the matrix model. 

As mentioned in Section~\ref{sec:T-duality}, this result is 
expected because the free energy of the dual model is the 
same as the original one (rigorously speaking, in the sense of the genus expansion) 
and hence it should satisfy 
the same string equation in the double scaling limit 
as the free energy of the original theory does, 
which then implies that both theory have the same form 
of the leading nonperturbative effects. Another argument which supports 
our result is that the leading part of the nonperturbative effect  
is given by the disk amplitude in the ZZ-brane background, 
but at the level of the disk topology the T-dual relation (\ref{identification}) holds. 
In this interpretation, it is crucial that the disk amplitudes $\vev{\tr \log(y\pm Y')}_d'$ 
cancel out in the process of the computation and do not contribute to the instanton effects 
as shown in Section~\ref{sec:nonperturbative effect}. 

However, regarding the next-to-leading part 
in the chemical potential of the instanton, the string equation does not guarantee
that the dual theory has the same coefficients 
in the nonperturbative effects as those in the original theory. 
In the direct calculation from the matrix model in Section~\ref{sec:coefficient}, 
several cylinder amplitudes appear to contribute to them. 
In particular, the amplitudes 
$\vev{(\tr \log(y-Y'))^2}_c'$ and $\vev{\tr\log(y-Y') \tr\log(y+Y')}_c'$ 
represent quite singular dual spin configurations and their counterparts 
in the original model do not appear in the calculation of the nonperturbative effects~\cite{Ishibashi:2005zf}. 
Thus, if the above amplitudes persist in contributing after taking the double scaling limit, 
there is a possibility that the T-duality does not hold at the level of the next-to-leading order. 
From this point of view, the comparison 
of the coefficients in the nonperturbative effects 
in both theories is quite interesting because it may 
possibly reflect the violation of the T-duality. 
On the other hand, if the nonperturbative effects computed 
from the original and dual matrix models coincide  
even including their coefficients, 
this may imply that the matrix model is more fundamental 
than the string equation and that 
the nonperturbative effect in string theory has large universality  
which contains the T-duality. 
In fact, it is shown in \cite{Ishibashi:2005zf} that 
the coefficient in the nonperturbative effect is universal 
within the original $c<1$ string theory in the sense that 
it does not depend on details of the potential of the matrix model. 
Therefore, it is an interesting problem to investigate whether 
this universality involves even the T-duality. 
We hope that this kind of study would give some insight into universality 
of string theory itself. For this purpose, it is desirable to calculate 
the cylinder amplitudes of trace-log operators 
like (\ref{trlog1}) and (\ref{trlog2}) 
which contribute to the coefficients. 
We hope we will be able to report on it in the near future.    

Apart from the coefficients of the nonperturbative effects, 
there are lots of quantities of interest to be computed. 
For example, 
in order to reinforce the identification between the dual ZZ-brane 
and the instanton in the dual matrix model, it is preferable 
to calculate the loop amplitudes under the corresponding backgrounds for both 
of the continuum theory and the matrix model 
and then to compare the obtained results.  
In particular, it is quite interesting 
how the T-duality and the dual ZZ-branes are realized 
in the continuum Liouville theory. 
In the case of the standard $c=1/2$ conformal field theory, 
we know how the T-duality maps the relevant operators into themselves. 
If we clarify it for the gravitationally dressed case 
($c=1/2$ noncritical string theory), we will be able to make 
more explicit the correspondence between relevant operators 
and associated ZZ-branes in the dual theory and  to construct 
the T-duality transformation law of the ZZ-branes 
in the noncritical string theory. The result in this paper strongly suggests 
that the T-duality transformation rule of the relevant operators
is the same as in the nongravitational case and that the T-duality does not change the tension 
of the ZZ-brane associated with each relevant operator.   

Finally, in refs.~\cite{Sugino:1995hk} and \cite{Asatani:1996jc}, $c=1/2$ noncritical string field theories 
corresponding to the original and dual two-matrix models have been constructed respectively, and it is 
clarified how string fields in the two theories are related each other under the T-duality. 
It is quite interesting to express the original and dual ZZ-branes in terms of the string fields\footnote
{For somewhat related works, see refs.~\cite{Fukuma:2006qq}.
}
and to determine the T-duality transformation rule between the ZZ-branes 
by utilizing the established transformation rule of the string fields. 
It will be also intriguing from the viewpoint of finding the relations 
between different descriptions of nonperturbative 
objects in string theory -- (ZZ-)branes and string fields.

\section*{Acknowledgements}
We would like to thank I. Kostov, N. Orantin and A. Yamaguchi 
for invaluable discussions. 
This work is benefited by the SAKURA project exchanging researchers between France and Japan. 
Also, one of the authors (F. S.) thanks the members of theoretical physics laboratory of RIKEN 
for hospitality during his stay, when a part of this work was done.

\appendix
\section{Double scaling limit and Resolvents in the dual model}
\label{app:dsl}
\setcounter{equation}{0}
In the paper \cite{Asatani:1996jc}, various amplitudes in the dual two-matrix model (\ref{dual}) 
are calculated and their expressions in the continuum limit are obtained. 
Here, for convenience, we present the result of the disk amplitude for the resolvent operator of $X$:
\be
R_X(x) = \vev{\frac{1}{N}\tr \frac{1}{x-X}}_d 
\ee
derived in \cite{Asatani:1996jc}. 
Also, we compute the disk amplitude for the resolvent of $Y$: 
\be
R_Y(y) = \vev{\frac{1}{N}\tr \frac{1}{y-Y}}_d, 
\ee
which has not appeared in the literature. 

Since the dual model (\ref{dual}) is obtained from the original one (\ref{original}) 
by changing the integration variables as (\ref{T-dual}), 
the partition functions for both models coincide and have the same critical point. 
The original model has the critical point
\be
c_* = \frac{-1+2\sqrt{7}}{27}, \quad g_*=\sqrt{10c_*^3}, 
\ee
and the double scaling limit to the $c=1/2$ string theory is taken as 
$N\to\infty$ and the lattice spacing $a\to 0$ with fixing 
\be
c=c_*, \quad t = \frac{g_* -g}{g_*a^2} \quad {\rm and} \quad \frac{1}{g_s}= ({\rm const}) N a^\frac{7}{3}. 
\label{dsl_original}
\ee
$t$ and  $g_s$ represent the cosmological constant and the string coupling constant in the continuum theory. 
Correspondingly, the critical point of the dual model is 
\be
(c, \hg) = (c_*,\hg_*)=\left(\frac{-1+2\sqrt{7}}{27}, \sqrt{5c_*^3}\right), 
\ee
and we obtain the dual $c=1/2$ string theory by taking the double scaling limit 
$N\to\infty$, $a\to 0$ with keeping 
\be
c=c_*, \quad t = \frac{\hg_* -\hg}{\hg_*a^2} \quad {\rm and} \quad \frac{1}{g_s}= ({\rm const}) N a^\frac{7}{3}
\label{dsl_dual}
\ee
fixed.  
In what follows, the coupling $c$ is understood to be fixed at the critical point $c_*$.

\subsection{$R_X(x)$}
\label{sec:RX}
As in \cite{Asatani:1996jc,Carroll:1995nj}, the following closed equation for $R_X(x)$ is derived by combining 
five Schwinger-Dyson equations obtained in the planar limit: 
\be
\hg R_X(x)^3 + f_2 R_X(x)^2 + f_1 R_X(x) + f_0 = 0, 
\label{SD_RX}
\ee
where 
\bea
f_2 & = & \hg^2x^2 + (5c-1)\hg x -2c(1+c), \nono \\
f_1 & = & 4c\hg^2x^3 - (6c-2c^2) \hg x^2 + 2c(1-c^2)x - 2\hg^2 V_X + \hg (1-3c), \nono \\
f_0 & = & (6c-2c^2-4c\hg x)\hg V_X - 4c\hg^2 V_{X^2} -4c\hg^2x^2 \nono \\
    &   & + (6c-2c^2)\hg x -2c(1-c^2) -\hg^2.
\eea
We used the notation: $V_{X^nY^m}\equiv\vev{\frac{1}{N}\tr (X^nY^m)}_d$. 
The amplitudes $V_{X^n}$ ($n = 1, 2$) are related to the amplitudes 
in the original model $W_n \equiv \vev{\frac{1}{N}\tr \,A^n}_d$ ($n = 1, 3$) as 
\be
V_X = \sqrt{2} W_1, \quad 
V_{X^2} = \frac{1-c^2}{cg}W_1 - \frac{g}{c}W_3 -\frac{1}{c}. 
\ee
After the shift 
\be
R_X(x) = -\frac{f_2}{3\hg} + \hat{R}_X(x), 
\label{shift_RX}
\ee
(\ref{SD_RX}) becomes 
\be
\hat{R}_X(x)^3 -\frac13F_1 \hat{R}_X(x)-\frac{1}{27}F_0=0
\label{SD_RX_2}
\ee
with
\be
F_1 = \frac{1}{\hg^2}f_2^2 -\frac{3}{\hg}f_1, \quad
F_0 = \frac{9}{\hg^2}f_1f_2 - \frac{2}{\hg^3}f_2^3 -\frac{27}{\hg}f_0. 
\ee
At $\hg=\hg_*$, $F_1$ and $F_0$ can be written as 
\be
F_1 = \frac{c}{5}(z-\hat{s})^4, \quad 
F_0 = -2\left(\frac{c}{5}\right)^{\frac32}(z-\hat{s})^4\left(z-\hat{s}+3\sqrt{6}\right)\left(z-\hat{s}-3\sqrt{6}\right), 
\label{F0F1*}
\ee
where $z\equiv \sqrt{5c}\, x$, $\hat{s}=1+\sqrt{7}$, and the fact was used that $W_1$, $W_3$ 
expressed in \cite{Sugino:1995hk}: 
\bea
W_1 & = & W_1^{\rm non} + \hat{W}_1, 
\label{W1} \\
W_1^{\rm non} & = & \frac{-8\rho_*^4+3(2g_*)^2}{3(2g_*)^3} + a^2\frac{-136\rho_*^4+27(2g_*)^2}{27(2g_*)^3}t, \nono \\
\hat{W}_1 & = & a^{\frac83}\frac{8\rho_*^4}{27(2g_*)^3}(5t)^{\frac43}+a^\frac{10}{3}\frac{4\rho_*^4}{81(2g_*)^3}(5t)^\frac{5}{3} 
+ a^4\frac{-8527\rho_*^4+972(2g_*)^2}{972(2g_*)^3}t^2 + \cO\left(a^\frac{14}{3}\right), \nono \\
W_3 & = & W_3^{\rm non} + \hat{W}_3, 
\label{W3} \\
W_3^{\rm non} & = & \frac{32(420-839\rho_*)\rho_*^5}{729(2g_*)^5} + a^2\frac{160(252-611\rho_*)\rho_*^5}{729(2g_*)^5}t, \nono \\
\hat{W}_3 & = & a^{\frac83}\frac{320\rho_*^6}{81(2g_*)^5}(5t)^{\frac43}+ a^\frac{10}{3}\frac{160\rho_*^6}{243(2g_*)^5}(5t)^\frac{5}{3} 
+ a^4\frac{70(1152-3593\rho_*)\rho_*^5}{729(2g_*)^5}t^2 + \cO\left(a^\frac{14}{3}\right), \nono
\eea
with $\rho_* = 3c$ 
become at the critical point 
\bea
W_{1*} & = & \left(\frac{9}{40c}\right)^{\frac32}\left(-8c+\frac{40}{27}\right), \nono\\
W_{3*} & = & \frac{32}{243}\left(\frac{9}{40c}\right)^{\frac52}\left(140-839c\right). \nono
\eea
Eqs. (\ref{F0F1*}) determine the critical point of $x$, denoted by $\hat{P}_*'$, as 
\be
\hat{P}_*' = \frac{\hat{s}}{\sqrt{5c}}, 
\ee
where the cubic equation (\ref{SD_RX_2}) becomes triply degenerate: $\hat{R}_X(\hat{P}_*')^3=0$. 
In general, the equation (\ref{SD_RX_2}) is solved by 
\be
\hat{R}_X(x) = \frac13\left[ \left(\frac{F_0+\sqrt{F_0^2-4F_1^3}}{2}\right)^{\frac13} + 
\left(\frac{F_0-\sqrt{F_0^2-4F_1^3}}{2}\right)^{\frac13} \right]. 
\label{sol_SD_RX}
\ee
Introducing the variables in the continuum theory $\hat{T}$ and $\zeta$ as 
\be
\hg = \hg_*\left(1-a^2\frac{\hat{s}^2}{10}\hat{T}\right), \quad x=\hat{P}_*'(1+a\zeta)
\ee
($\hat{T}$ is proportional to the cosmological constant $t$), the continuum limit of the solution 
(\ref{sol_SD_RX}) takes the form 
\be
\hat{R}_X(x) =a^{\frac43}\hat{\alpha}'w(\zeta)+ \cO\left(a^{\frac53}\right) , \quad 
w(\zeta)= \huniv{\zeta}, 
\label{RX_cont}
\ee
with $\hat{\alpha}'=\frac{c^{\frac12}\hat{s}^{\frac43}}{2^{\frac23}\sqrt{5}}$. 
Also, 
\be
R_X(x) = \Rnon_X(x) + \hat{R}_X(x), \quad
\Rnon_X(x) = -\frac{f_2}{3\hat{g}}, 
\label{RXnon}
\ee
where $\Rnon_X$ and $\hat{R}_X$ stand for the nonuniversal and universal parts, respectively.                            

Eigenvalue distribution of the matrix $X$ is seen as a cut of $\hat{R}_X(x)$ on the $x$-plane, 
which is determined 
in principle from (\ref{sol_SD_RX}) under the condition of $\hat{R}_X(x)$ having one cut and being analytic for 
$\Rep \,x > x_c$ with $x_c$ some finite number. 
However, practically it is not an easy task in general, 
since $F_0^2-4F_1^3$ is a complicated polynomial of $x$ with the degree ten. 
$w(\zeta)$ in (\ref{RX_cont}) has one cut $\zeta\in \left(-\infty, -\sqrt{\hat{T}}\right]$, 
which means that the right edge of the cut for $\hat{R}_X(x)$ is $x=\beta'\equiv \hat{P}_*'\left(1-a\sqrt{\hT}\right)$ 
when $\hg$ is near the critical point $\hat{g}_*$. 
The leading value of the left edge also can be determined from the expression of (\ref{F0F1*}). 
Plugging it into (\ref{sol_SD_RX}), we find the form of $\hR_X(x)$, with $c$ and $\hg$ set to the critical values 
but $x$ left generic, as 
\bea
\hR_X(x) &= &\frac13\sqrt{\frac{c}{5}} \, (z-\hat{s})^{\frac43}\left(g_+(z)^{\frac13} + g_-(z)^{\frac13}\right), 
\label{Runiv_cr}\\
g_{\pm}(z) & \equiv & -\left(z-\hat{s}+3\sqrt{6}\right)\left(z-\hat{s}-3\sqrt{6}\right)\pm 6\sqrt{3}
        \sqrt{-\left(z-\hat{s}+3\sqrt{3}\right)\left(z-\hat{s}-3\sqrt{3}\right)}. \nono
\eea
Here, it is easy to see that $g_+(z)$ has no zero and $g_-(z)$ has the unique zero at $z=\hat{s}$. 
For $z\sim \hat{s}$, 
\be
g_+(z) = 108 + \cO((z-\hat{s})^2), \quad 
g_-(z) = \frac{1}{108}(z-\hat{s})^4 + \cO((z-\hat{s})^5). 
\ee
{}From these properties of $g_\pm(z)$, it can be shown 
that (\ref{Runiv_cr}) has one cut of the interval $z\in [\hat{s}-3\sqrt{3}, \hat{s}]$. 
Thus, the left edge of the cut for $\hR_X(x)$ is given by 
$x= \alpha'\equiv \hP'-\frac{3\sqrt{3}}{\sqrt{5c}}+\cO(a)$ in the case of  $\hg$ near the critical point.

\subsection{$R_Y(y)$}
\label{sec:RY}
To obtain a closed equation for $R_Y(y)$, 
we combine the six Schwinger-Dyson equations derived from the following identities in the 
planar limit:
\bea
0 & = & \int d^{N^2}Xd^{N^2}Y \sum_\alpha\frac{\partial}{\partial Y_\alpha}\tr\left(\frac{1}{y-Y}t^\alpha\right)
            e^{-N\tr\, \tilde{S}_D(X,Y)}, \nono \\
0 & = & \int d^{N^2}Xd^{N^2}Y \sum_\alpha\frac{\partial}{\partial X_\alpha}\tr\left(\frac{1}{y-Y}t^\alpha\right)
             e^{-N\tr \, \tilde{S}_D(X,Y)}, \nono \\
0 & = & \int d^{N^2}Xd^{N^2}Y \sum_\alpha\frac{\partial}{\partial Y_\alpha}\tr\left(\frac{1}{y-Y}Xt^\alpha\right)
             e^{-N\tr \, \tilde{S}_D(X,Y)}, \nono \\
0 & = & \int d^{N^2}Xd^{N^2}Y \sum_\alpha\frac{\partial}{\partial Y_\alpha}\tr\left(\frac{1}{y-Y}Xt^\alpha X\right)
             e^{-N\tr \, \tilde{S}_D(X,Y)}, \nono \\
0 & = & \int d^{N^2}Xd^{N^2}Y \sum_\alpha\frac{\partial}{\partial X_\alpha}\tr\left(\frac{1}{y-Y}XYt^\alpha\right)
             e^{-N\tr \, \tilde{S}_D(X,Y)}, \nono \\
0 & = & \int d^{N^2}Xd^{N^2}Y \sum_\alpha\frac{\partial}{\partial X_\alpha}\tr\left(\frac{1}{y-Y}t^\alpha YX\right)
             e^{-N\tr \, \tilde{S}_D(X,Y)}. 
\eea
Here, $t^\alpha$ ($\alpha = 1, \cdots, N^2$) are a basis of $N\times N$ Hermitian matrices satisfying 
\be
\sum_\alpha \tr (Wt^\alpha Zt^\alpha) = \tr\, W \, \tr \, Z, \quad 
\sum_\alpha \tr (W t^\alpha)\, \tr (Zt^\alpha) = \tr (WZ)
\ee
for arbitrary matrices $W$, $Z$. 
The result is 
\be
R_Y(y)^4 + b_1 R_Y(y)^3 + b_2 R_Y(y)^2 + b_3 R_Y(y) + b_4 =0, 
\label{SD_RY}
\ee
where 
\bea
b_1 & = & -8cy, \nono \\
b_2 & = & 4\hg^2 y^4 + (-1+2c+19c^2) y^2 + 2(1+c-2\hg V_X), \nono \\
b_3 & = & -16c\hg^2 y^5 + 4c(1+c)(1-3c) y^3 -8c(1+c -2\hg V_X) y , \nono \\
b_4 & = & 16c\hg^2 y^4 \nono \\
    &   & + \{-4c(1+c)(1-3c)-4(1-c)(1-3c)\hg V_X + 4(1-3c)\hg^2 V_{X^2} + 8\hg^3 V_{XY^2}\} y^2 \nono \\
    &   & +(1+c-2\hg V_X)^2. 
\eea
(Note that $V_{Y}=V_{XY}=0$ from the $\mathbf{Z}_2$ symmetry under $Y\to -Y$.) 
Also, $V_{XY^2}$ can be expressed in terms of the amplitudes $W_1$ and $W_3$ in the original model, 
similarly to $V_X$ and $V_{X^2}$, as 
\be
V_{XY^2} = \frac{1}{\sqrt{2}}\left(-\frac{(1-c)(1-c^2)}{cg^2}W_1 + \frac{1+c}{c}W_3 + \frac{1}{cg}\right). 
\ee
We shift $R_Y(y)$ as 
\be
R_Y(y) = -\frac14 b_1 + \hR_Y(y), 
\ee
so that the cubic term of $\hR_Y(y)$ in (\ref{SD_RY}) vanishes, to obtain 
\be
\hR_Y(y)^4 +B_1 \hR_Y(y)^2 + B_2 \hR_Y(y) + B_3 = 0, 
\label{SD_RY2}
\ee
where 
\bea
B_1 & = & -\frac38 b_1^2 + b_2, \nono\\
B_2 & = & \frac18 b_1^3 - \frac12 b_1b_2 + b_3, \nono \\
B_3 & = & -\frac{3}{256}b_1^4 + \frac{1}{16}b_1^2 b_2 -\frac14 b_1 b_3 + b_4. 
\eea
The critical points of $y$ are determined by 
\be
\left. B_1\right|_* = \left. B_2\right|_* = \left. B_3\right|_* = 0, 
\ee
with the symbol ``$|_*$'' meaning $c$, $\hg$, $W_1$ and $W_3$ set to the critical values. 
The equation for $B_2$ is trivially satisfied, and 
the remaining two lead to 
\bea
\left. B_1\right|_*  & = & 20c^3 (y-\hQ')^2(y+\hQ')^2 = 0, \nono \\
\left. B_3\right|_*  & = & -80c^5 (y-\hQ')^3(y+\hQ')^3 =0, 
\quad \left(\hQ' \equiv \frac{2}{\sqrt{5c}}\right), 
\label{B1B3*}
\eea
to determine the critical points as 
\be
y = \pm \hQ'.
\label{Qhat*}
\ee
The result reflects the $\mathbf{Z}_2$ symmetry as should be.   
We introduce the scaling variable $\eta$ as $y = \hQ' (1+a\eta)$ 
and take the continuum limit of the quartic equation (\ref{SD_RY2}) 
by using the expressions of $W_1$, $W_3$ 
in (\ref{W1}) and (\ref{W3}).   
The result is 
\be
\hR_Y(y)^4 -a^2\, \frac85 c (32\eta^2- \hat{s}^2\hT)\hR_Y(y)^2 
-a^3\, 2\left(\frac{8^2}{5}c\right)^2\eta^3 + \cO \left(a^{\frac{11}{3}}\right)=0, 
\label{SD_RY3}
\ee
which is solved as 
\bea
R_Y(y) & = & \Rnon_Y(y) + \hR_Y(y), \nono\\
\Rnon_Y(y) & = & -\frac14 b_1 = 2cy, \nono \\
\hR_Y(y) & = & a^{\frac34}\, 8\cdot 2^{\frac14} \sqrt{\frac{c}{5}} \, \eta^{\frac34} + \cO \left(a^{\frac54}\right). 
\label{sol_RY}
\eea
(We took the branch of $\hR_Y(y)$ to be real positive for $y$ or $\eta$ large positive.) 
The scaling of $\hR_Y(y)$ with the power $a^{\frac34}$ 
is clearly different from the scaling of correlators ever computed 
in the original and dual models \cite{Asatani:1996jc,Sugino:1995hk}, which is characterized by the power of cubic roots. 
$R_Y(y)$ is regarded as a disk amplitude with the disorder operators distributed densely over the boundary, 
and such an amplitude has not been computed in the literature concerning either of the original and dual theories.   
Also, the leading expression of $\hR_Y(y)$ does not depend on the cosmological constant, meaning that 
the disk collapses into a surface of zero area although its boundary has a finite length. 
In the above, we took the continuum limit around the critical point $y=\hQ'$.  
Similarly, the limit around the other critical point setting $y = -\hQ' (1+a\eta)$ leads to the same equation 
as (\ref{SD_RY3}), which is a consequence from the $\mathbf{Z}_2$ symmetry. 

The expression ({\ref{sol_RY}) has one cut $\eta\in (-\infty, 0]$, whose end point does not depend on $\hT$. 
It shows that 
the right edge of the cut $y=\hQ'$ on $y$-plane receives no $\cO(a)$ correction 
even when $\hg$ moves slightly off the critical point. 
In order to see the global structure of the cut on $y$-plane, we solve eq. (\ref{SD_RY2}) by using (\ref{B1B3*}) 
in the case of $(c, \hg)$ fixed at the critical point but $y$ left generic. 
The result is 
\be
\hR_Y(y) = \sqrt{10c^3}\, (y-\hQ')^{\frac34}(y+\hQ')^{\frac34}\left[y-\sqrt{(y-\hQ')(y+\hQ')}\right]^{\frac12}, 
\ee
(We took the same choice of the branch as above.) 
which has one cut of the interval $y\in [-\hQ', \hQ']$. 
Therefore, we see that $\hR_Y(y)$ has the $\mathbf{Z}_2$-symmetric one cut of the form 
$y\in [-\gamma', \gamma']$ with $\gamma' = \hQ' + o(a)$ for the case $\hg$ near the critical point.

\subsection{Expressions for the action (\ref{dual2})}  
The calculation of the instanton effects in Section~\ref{sec:nonperturbative effect} 
is done by using the rescaled action $S_D(X,Y)$ in (\ref{dual2}) instead of 
$\tilde{S}_D(X,Y)$ in (\ref{dual}). 
The former is obtained from the latter by 
\be
X \to \frac{1}{\hg} \, X, \quad Y \to \frac{1}{\hg} \, Y. 
\label{rescaling}
\ee
Here, we consider the effect of the rescaling to the results obtained in Sections~\ref{sec:RX} and \ref{sec:RY}, 
to match them with the expressions in Section~\ref{sec:nonperturbative effect}. 

Let us rename the resolvents calculated using the action $\tilde{S}_D(X,Y)$ as $\tilde{R}_X(x)$ and $\tilde{R}_Y(y)$, 
which are nothing but those appearing in the sections~\ref{sec:RX} and \ref{sec:RY}. 
As a result of the rescaling, the resolvents $R_X$, $R_Y$ considered in Section~\ref{sec:nonperturbative effect} 
are related to $\tilde{R}_X$, $\tilde{R}_Y$ as 
\be
R_X(x) = \frac{1}{\hg} \, \tilde{R}_X\left(\frac{1}{\hg}\, x\right), \quad 
R_Y(y) = \frac{1}{\hg} \, \tilde{R}_Y\left(\frac{1}{\hg}\, y\right). 
\ee
Hence, the critical points of $x$ and $y$ for $R_X(x)$ and $R_Y(y)$, denoted by $\hP$ and $\hQ$ respectively, 
are given as 
\be
\hP =  \hg_* \hP' = \hat{s}c=\frac{1+c}{2}, \qquad \hQ = \hg_*\hQ' = 2c. 
\ee
For $R_X(x)= \Rnon_X(x) + \hR_X(x)$, the nonuniversal part becomes 
\bea
\Rnon_X(x) & = & \frac{1}{\hg}\tilde{R}^{\rm non}_X\left(\frac{1}{\hg}x\right) 
                           = \left.-\frac{1}{3\hg^2}f_2\right|_{x \to \frac{1}{\hg}x} \nono \\
      & = & -\frac{1}{3h}(x^2 + (5c-1)x -2c(1+c)) \nono \\
     & = & \frac{1}{3h}(2V'(x) - V'(1+c-x)), 
\label{RXnon_final}     
\eea
where $h=\hg^2$, and $V(x) \equiv \frac{1-c}{2}x^2-\frac13x^3$. 
The scaling variable $\zeta$ is introduced as $x=\hP (1+a\zeta)$, and the universal part 
takes the form 
\bea 
\hR_X(x) & = & \frac{1}{\hg}\hat{\tilde{R}}_X\left(\frac{1}{\hg}x\right) 
                   = a^{\frac43}\hat{\alpha} w(\zeta) + \cO \left(a^{\frac53}\right), \nono\\
\ha & = & \frac{1}{\hg_*}\ha' = \frac{\hat{s}^\frac{4}{3}}{2^\frac{2}{3}\cdot 5c}, \nono\\
w(\zeta) & = & \huniv{\zeta}. 
\label{RXuniv_final} 
\eea
When $\hg$ is near the critical point, endpoints of the cut of $\hR_X(x)$: $x\in I_X= [\alpha, \beta]$ are 
expressed as 
\bea
\beta & = & \hg_*\beta' = \hP \left(1-a\sqrt{\hT}\right), \nono\\
\alpha & = & \hg_* \alpha' = \hP -3\sqrt{3}c + \cO(a). 
\label{cut_IX}
\eea
Similarly, for $R_Y(y)= \Rnon_Y(y) + \hR_Y(y)$, we define the variable $\eta$ by $y=\hQ (1+a\eta)$, 
and obtain 
\bea
\Rnon_Y(y) & = & \frac{1}{\hg}\tilde{R}^{\rm non}_Y\left(\frac{1}{\hg}y\right) 
      =  \left. -\frac{1}{4\hg}b_1\right|_{y \to \frac{1}{\hg}y}
      = -\frac{2c}{h}y, \nono \\
\hR_Y(y) & = & \frac{1}{\hg}\hat{\tilde{R}}_Y\left(\frac{1}{\hg}y\right) 
       = a^{\frac34} \, \frac{8\cdot 2^{\frac14}}{5c}\eta^{\frac34} + \cO\left(a^{\frac54}\right). 
\label{RY_final}
\eea
The cut of $\hR_Y(y)$ is given by the $\mathbf{Z}_2$-symmetric interval $y\in I_Y$ as 
\be
I_Y = [-\gamma, \gamma] \quad {\rm with} \quad \gamma = \hQ + o(a), 
\label{cut_IY}
\ee
for the case $\hg$ near the critical point.  

Since the one-matrix model (\ref{Z_O(1)}) is obtained after the Gaussian integration over $Y$ 
in (\ref{dual2}), the partition functions of (\ref{dual2}) and (\ref{Z_O(1)}) are identical and have 
the same double scaling limit (\ref{dsl_dual}). Also, concerning the correlators among operators 
independent of $Y$ (for example, $R_X(x)$), both models give the same result.

\section{Computation of denominator}
\label{app:denominator}
\setcounter{equation}{0}
In this appendix, we compute the denominator in (\ref{mu}) 
based on the method developed in \cite{Ishibashi:2005dh}. 
Using (\ref{Z^0}), the denominator can be written in terms of 
the partition functions as 
\be
\int_{(x,y)\in\cS} dxdy \,e^{-\Veff (x,y)}
 = \frac{D_NZ_N^{\rm (0-inst)}(h)}{D_{N-1}Z_{N-1}(h')}.
\label{den.def} 
\ee
The difference between the 0-instanton partition function 
$Z_N^{\rm (0-inst)}(h)$ and the total partition function $Z_N(h)$ is 
exponentially small as $\exp(-C/g_s)$ which is nothing but the 
nonperturbative effect. Since it is negligible in the computation of $\mu$, 
we can replace $Z_N^{\rm (0-inst)}(h)$ with $Z_N(h)$ 
in (\ref{den.def}). Therefore, 
\be
\int_{(x,y)\in\cS} dxdy \,e^{-\Veff (x,y)}
 = \frac{D_NZ_N(h)}{D_{N-1}Z_{N-1}(h')}.
\ee
Namely, the denominator can be obtained basically as the ratio
between the partition functions of the matrix model with rank $N$ 
and that with rank $N-1$. As mentioned in (\ref{Fexp}), 
under the measure (\ref{measure}) 
the total free energy has the standard $1/N$-expansion 
\bea
Z_N(h) & = & \exp\left[-N^2F_0(h)-F_1(h)+\cO\left(\frac{1}{N^2}\right)\right], \nono\\
Z_{N-1}(h') & = & \exp\left[-(N-1)^2F_0(h')-F_1(h')
                                   +\cO\left(\frac{1}{(N-1)^2}\right)\right].
\eea
Hence, 
\bea
\lefteqn{\int_{(x,y)\in\cS} dxdy \,e^{-\Veff (x,y)}}\nonumber \\
&=&\frac{D_N}{D_{N-1}}
     \exp
     \left[-(N-1)(2F_0(h')+h'F'_0(h'))
          -\left(F_0(h')+2h'F'_0(h')
            +\frac{1}{2}h^{\prime 2}F_0^{\prime\prime}(h')\right)\right] 
            \nonumber \\
 & & \times \left[1+\cO\left(\frac{1}{N}\right)\right].                
\label{denominator}
\eea
The computation is essentially reduced to finding the sphere free energy $F_0(h)$. 

\subsection{Sphere free energy}
In order to compute the sphere free energy, 
we first perform the integration with respect to $Y$ in (\ref{dual2}) 
and express $Z_N(h)$ as the integration over 
the eigenvalues of $X$. Then we obtain 
\begin{align}
Z_N(h)
 & = &  {J_N^X}^{-1}\int \prod_{i=1}^Nd\lam_i 
       \exp\Biggl[&\sum_{i<j}\log (\lam_i-\lam_j)^2
                 -\prefI\sum_iV(\lam_i) \nono \\
 &&-\frac{1}{2}&\sum_{ij}
                  \log\left(c+1-\lam_i-\lam_j\right)     
                 +\frac{N^2}{2}\log (c+1)\Biggr], 
\label{anotherZ}                                
\end{align}
where $J_N^X$ is defined in (\ref{J^X}) whose explicit form is given 
in Appendix~\ref{sec:DJ}. In the large-$N$ limit, by introducing the eigenvalue distribution 
$\rho(\lambda)=\vev{\frac{1}{N}\sum_i\delta (\lambda-\lambda_i)}$, 
this becomes 
\begin{align}
Z_N^{(0)}(h)  =  
{J_N^X}^{-1}\exp
\Biggl[N^2\Biggl\{&\int d\lam d\mu \rho(\lam)\rho(\mu)\log|\lam-\mu|
                -\frac{1}{h}\int d\lam \rho(\lam)V(\lam)\nono \\
    -\frac{1}{2}&\int d\lam d\mu \rho(\lam)\rho(\mu)
                           \log\left(c+1-\lam-\mu\right)
    +\frac{1}{2}\log (c+1)\Biggr\}\Biggr]. 
    \label{Zbyrho}
\end{align}
Therefore, the sphere free energy is expressed in terms of the eigenvalue 
distribution as 
\begin{align}
F_0(h) = -&\int d\lam d\mu \rho(\lam)\rho(\mu)\log |\lam-\mu|
        +\frac{1}{h}\int d\lam\rho(\lam)V(\lam) \nono \\
        +\frac{1}{2}&\int d\lam d\mu \rho(\lam)\rho(\mu)
              \log(c+1-\lam-\mu) 
        -\frac{1}{2}\log (c+1)+\frac{1}{N^2}\log J_N^X.
        \label{F_0}       
\end{align}

\subsection{Identity of the resolvent}
Here we derive an identity of the resolvent $R_X(z)$ 
introduced in (\ref{resolvent}) which plays an important role 
in the following. 
In terms of the eigenvalue distribution $\rho(\lambda)$, 
the resolvent is given by 
\be
R_X(z)=\int d\lambda\frac{\rho(\lambda)}{z-\lambda}.
\label{rho&R}
\ee
Then the support of the eigenvalue distribution is characterized 
as the cut of $R_X(z)$ on the complex $z$-plane. 

On the other hand, in the large-$N$ limit $\rho(\lambda)$ 
satisfies a saddle point equation which follows from 
(\ref{F_0}) 
\bea
0 & = & 2P\int d\mu \frac{\rho(\mu)}{\lam-\mu}-\frac{1}{h}V'(\lam)
         +\int d\mu \frac{\rho(\mu)}{c+1-(\lam+\mu)} \nono \\
  & = & R_X(\lam+i0)+R_X(\lam-i0)+R_X(c+1-\lam)-\frac{1}{h}V'(\lam),
  ~~~{\rm for}~\lam~{\rm on~the~cut}.\nono \\
\label{SPE}
\eea
Note that, for $\lambda$ on the cut, $\lambda<\hP$ as shown in 
(\ref{cut_IX}) and $c+1-\lambda>\hP$ is always outside the cut. 
Hence the third term in the last equation in (\ref{SPE}) does not 
have the imaginary part. A particular solution to this equation 
is provided by 
\be
R_X^{\rm non}(z)=\frac{1}{3h}(2V'(z)-V'(1+c-z)).
\label{Rnon}
\ee
Therefore, the resolvent $R_X(z)$ is in general given by 
\be
R_X(z)=\Rnon_X(z)+\hR_X(z),
\label{decomp of R_X}
\ee
where $\hR_X(z)$ satisfies 
\be
0=\hR_X(z+i0)+\hR_X(z-i0)+\hR_X(1+c-z),
~~~{\rm for}~z~{\rm on~the~cut}. 
\label{homogeneousSPE}
\ee
In fact, this form of the resolvent coincides with the result 
in (\ref{RXnon_final}) and (\ref{RXuniv_final}) 
in Appendix \ref{app:dsl}.  

\subsection{$F_0(h)$ in terms of the resolvent}
Now let us express each term in (\ref{F_0}) in terms of the resolvent. 
{}From (\ref{rho&R}), it is easy to see that the following identities 
hold: 
\bea
\lefteqn{\int dzdz'\rho(z)\rho(z')\log|z-z'|} \nono \\
 & = & \int_{\Lambda}^{\beta}R_X(z')dz'+\log\Lambda
      +\int_{\alpha}^{\beta}dz\rho(z)\int_{\beta}^z\Rep R_X(z')dz',
      \nono\\
\lefteqn{\int dzdz'\rho(z)\rho(z')\log(c+1-z-z')} \nono\\
 & = & 
 \int_{\Lambda}^{\beta}R_X(z')dz'+\log\Lambda
      +\int_{\alpha}^{\beta}dz\rho(z)
       \int_{\beta}^{c+1-z}R_X(z')dz', 
\eea
where $\Lambda\rightarrow\infty$ limit is understood 
to be taken eventually, and $\beta$ is the right edge of the cut 
explicitly given in (\ref{cut_IX}). 
Substituting these into (\ref{F_0}) 
and using the saddle point equation (\ref{SPE}) leads to 
\be
F_0(h)=-\frac{1}{2}\vev{\frac{1}{N}\tr\log(\beta-X)}_d
      -\frac{1}{2}\int_{c+1-\beta}^{\beta}R_X(z')dz'
      +\frac{1}{2h}V(\beta)
      +\frac{1}{2h}\int dz\rho(z)V(z)+\cJ,
\label{F_0final}      
\ee
where $\cJ$ represents terms independent of the potential $V$
\be
\cJ=-\frac{1}{2}\log(c+1)+\frac{1}{N^2}\log J_N^X.
\label{defofcJ}
\ee
According to (\ref{denominator}), we also need derivatives of $F_0(h)$ 
in computing the denominator. In order to make explicit 
$h$-dependence of the sphere free energy, 
we rewrite $F_0(h)$ again in terms of the integration with respect to  
the eigenvalues as 
\bea
\lefteqn{F_0(h)-\cJ}\nono \\ 
& = & -\frac{1}{N^2}\log
      \left[\int \prod_{i=1}^N d\lam_i \, \Delta^{(N)}(\lam)^2
      e^{-\prefI\sum_iV(\lam_i)}
      \left(\frac{1}{1+c}\right)^{\frac{N^2}{2}}
      \prod_{i,j=1}^N
      \left(1-\frac{1}{1+c}(\lam_i+\lam_j)\right)^{-\frac{1}{2}}\right],                     
      \nonumber\\
\eea
where we have used (\ref{anotherZ}). This shows 
\bea
h\frac{\der}{\der h}(F_0(h)-\cJ)
=-\frac{1}{N^2}\vev{\prefI \tr V(X)}_d=-\frac{1}{h}\int dz \rho(z)V(z).
\label{F_0der}
\eea
{}From (\ref{F_0final}) and (\ref{F_0der}), 
for the $(N-1)\times (N-1)$ 
matrix model with $h'$ defined as (\ref{h'}), we obtain 
\bea
\lefteqn{2F_0(h')+h'F'_0(h')}\nono \\
 & = & -\vev{\frac{1}{N-1}\tr\log(\beta-X')}'_d
      -\int_{c+1-\beta}^{\beta}R_{X'}(z')dz'
      +\frac{1}{h'}V(\beta)+\left(2+h'\frac{\der}{\der h'}\right)\cJ',
      \nono \\
\label{denexpre}      
\eea
where $\cJ'$ is obtained by replacing $N$ and $h$ by $N-1$ and $h'$ 
in (\ref{defofcJ}). 
We will see below that the last term in this equation 
together with the overall factor $D_N/D_{N-1}$ in (\ref{denominator}) 
gives subleading contributions not affecting the leading term. Therefore,  
we find that the leading term in the denominator is given as 
\bea
\lefteqn{\left.\int_{(x,y)\in\cS} dxdy \,e^{-\Veff (x,y)}
\right|_{\text{leading}}}\nonumber\\
 & = & \exp\left[\vev{\tr\log(\beta-X')}'_d
      +(N-1)\int_{c+1-\beta}^{\beta}R_{X'}(x')dx'
      -\prefII V(\beta)\right]. 
\label{denfinal}      
\eea 

\subsection{Next-to-leading contributions in the denominator}
In this subsection, we evaluate next-to-leading contributions in the denominator 
in (\ref{mu}). 

\subsubsection{Derivation of $D_N$ and $J_N^X$}
\label{sec:DJ}
First let us evaluate the measure factors $D_N$, $J_N^X$ 
introduced in (\ref{Z}) and (\ref{J^X}). 

In order to calculate $J_N^X$, we make a connection 
between the measure $dX$ defined in (\ref{measure}) 
and the standard measure 
\be
\widetilde{dX}
\equiv \prod_idX_{ii}\prod_{i<j}2(d \, \Rep X_{ij})(d \, \Imp X_{ij}).
\ee
For this purpose, it is sufficient to consider the Gaussian integration as follows. 
In the case of the standard measure, 
\be
\int \widetilde{dX}e^{-\frac{1}{2}\tr X^2}=(2\pi)^{\frac{N^2}{2}}. 
\ee
Comparing this to (\ref{measure}) yields  
\be
dX=\left(\frac{(1-c)N}{2\pi h}\right)^{\frac{N^2}{2}}\widetilde{dX}. 
\ee
On the other hand, by using the method of orthogonal polynomials, 
\be
J_N^X\int dXe^{-\frac{1}{2}\tr X^2}=
\int \prod_id\lam_i \, \Delta^{(N)}(\lam)^2e^{-\frac{1}{2}\sum_i\lam_i^2}
=N!\prod_{n=0}^{N-1}h_n
=(2\pi)^\frac{N}{2}\prod_{p=0}^Np!,
\ee
where we have used the well-known fact that in the Gaussian case 
the orthogonal polynomial is nothing but the Hermite polynomial 
with $h_n=\sqrt{2\pi}n!$. Thus we find  
\be
J_N^X=(2\pi)^{\frac{N}{2}}\left(\prod_{p=0}^Np!\right)
       \left(\frac{h}{(1-c)N}\right)^{\frac{N^2}{2}}.
\label{J_N^X}       
\ee
Similarly to $J_N^X$ in (\ref{J^X}), if we introduce $J_N^Y$ by 
\be
J_N^Y\int dY f(\tr Y, \tr Y^2, \cdots)
=\int\left(\prod_{i=1}^{N}d\mu_i\right) \, \Delta^{(N)}(\mu)^2f\left(\sum_i\mu_i, \sum_i\mu_i^2, \cdots\right) 
\label{J_Ndef}
\ee
with $dY$ defined in (\ref{measure}), 
we obtain 
\be
J_N^Y=(2\pi)^{\frac{N}{2}}\left(\prod_{p=0}^Np!\right)
       \left(\frac{h}{(1+c)N}\right)^{\frac{N^2}{2}}.
       \label{J_N^Y}
\ee
Using these, (\ref{dual2}) becomes 
\bea
\lefteqn{\int dXdY e^{-\prefI 
\tr\left(V(X)+\frac{1+c}{2}Y^2-XY^2\right)}} \nono \\
& &=  (J_N^XJ_N^Y)^{-1}\int \left(\prod_{i=1}^{N}d\lam_id\mu_i\right) \, 
 \Delta^{(N)}(\lam)^2\Delta^{(N)}(\mu)^2 \,
 e^{-\prefI\sum_i\left(V(\lam_i)+\frac{1+c}{2}\mu_i^2\right)}
 I(\lam,\mu), \nono \\
\label{JXJY} 
\eea
where 
\be
I(\lam,\mu)=\int dU\exp\left(\prefI \tr(XUY^2U^{\dagger})\right).
\ee
It is calculated by the method in \cite{Itzykson:1979fi} as 
\be
I(\lam,\mu)=\left(\prefI \right)^{-\frac{N(N-1)}{2}}
\left(\prod_{p=1}^{N-1}p!\right) \, \frac{\det_{ij}
e^{\prefI \lam_i\mu_j^2}}{\Delta^{(N)}(\lam)\Delta^{(N)}(\mu^2)}.
\ee
Substituting this into (\ref{JXJY}), we find 
\bea
\lefteqn{\int dXdY e^{-\prefI 
\tr\left(V(X)+\frac{1+c}{2}Y^2-XY^2\right)}}\nono \\
&&=(J_N^XJ_N^Y)^{-1}\left(\prefI\right)^{-\frac{N(N-1)}{2}}
\left(\prod_{p=1}^{N}p!\right) \nono \\
& &\hspace{7mm} \times \int \left(\prod_{i=1}^{N}d\lam_id\mu_i\right) \, 
 \frac{\Delta^{(N)}(\lam)\Delta^{(N)}(\mu)}{\prod_{i<j}(\mu_i+\mu_j)} \, 
 e^{-\prefI\sum_i\left(V(\lam_i)+\frac{1+c}{2}\mu_i^2
    -\lam_i\mu^2_i\right)}.
\eea
Comparing this equation with the definition of $D_N$ (\ref{Z}), 
we finally obtain 
\be
D_N
=J_N^XJ_N^Y\left(\prefI\right)^{\frac{N(N-1)}{2}}
\left(\prod_{p=1}^{N}p!\right)^{-1}
=\left(\prod_{p=0}^Np!\right)\left(\frac{4\pi^2 h}{N}\right)^{\frac{N}{2}}
 \left(\frac{h}{(1-c^2)N}\right)^{\frac{N^2}{2}}. 
\label{D_Nexpre}
\ee
Thus the overall factor in (\ref{denominator}) becomes  
\bea
\frac{D_N}{D_{N-1}}
= (2\pi)^{\frac{3}{2}}\sqrt{N-1} \, e^{-(N-1)}
    \frac{{h'}^N}{(1-c^2)^{N-\frac{1}{2}}}
    \left(1+\cO\left(\frac{1}{N}\right)\right),         
\label{dencoefftemp}      
\eea
which agrees with the result in \cite{Ishibashi:2005zf}. 

\subsubsection{Other contributions in (\ref{denominator})}
Using the Euler-Maclaurin formula, we evaluate the following quantity
\be
\log \left(\prod_{p=0}^Np!\right)=\log \left(\prod_{k=1}^{N}k^{N+1-k}\right)
=\frac{(N+1)^2}{2}\log (N+1)-\frac{3}{4}N^2-\frac{N}{2}+\cO (1), 
\ee
which is also derived in \cite{Kawai:2004bz}. 
{}From (\ref{J_N^X}), we see that $\cJ$ given in (\ref{defofcJ}) 
becomes 
\begin{align}
\cJ =&-\frac{1}{2}\log (c+1)
      +\frac{1}{N^2}\log \left(\prod_{p=0}^{N}p!\right)
      +\frac{1}{2}\log\left(\frac{h}{(1-c)N}\right)     
      +\cO\left(\frac{1}{N}\log N\right) \nono \\
    =&-\frac{1}{2}\log\left(\frac{(1-c^2)N}{h}\right)
      +\frac{1}{N^2}
       \left(\frac{(N+1)^2}{2}\log (N+1)
            -\frac{3}{4}N^2\right)
      +\cO\left(\frac{1}{N}\log N\right) \nono \\
    =&-\frac{1}{2}\log\left(\frac{(1-c^2)}{h}\right)-\frac{3}{4}
      +\cO\left(\frac{1}{N}\log N\right).
\label{cJexpre}                
\end{align}
Therefore, the last term in (\ref{denexpre}) gives  
\be
\left(2+h'\frac{\der}{\der h'}\right)\cJ'
=-\log\left(\frac{1-c^2}{h'}\right)-1.
\label{sub1}
\ee
Next, let us examine the ${\cal O}(N^0)$ contribution in the exponent 
in (\ref{denominator}). From (\ref{F_0der}) we have 
\be
h'F'_0(h')
=-\frac{1}{h'}\int dz\rho(z)V(z)+\frac{1}{2}. 
\ee
Combining this equation with (\ref{denexpre}) and (\ref{sub1}), 
we find 
\be
F_0(h')+2h'F'_0(h')+\frac{1}{2}h^{\prime 2}F_0^{\prime\prime}(h')
=-R-h'\frac{\der}{\der h'}R
 -\frac{1}{2}\log\left(\frac{1-c^2}{h'}\right),
 \label{O(1)contribution}
\ee
where $R$ is given as 
\bea
R & \equiv & \frac{1}{2}\vev{\frac{1}{N-1}\tr\log(\beta-X')}'_d
            +\frac{1}{2}\int_{c+1-\beta}^{\beta}R_{X'}(z')dz'
            -\frac{1}{2h'}V(\beta). 
\label{R_def}            
\eea
Also, (\ref{denexpre}) is written as 
\be
2F_0(h')+h'F'_0(h')=-2R-\log\left(\frac{1-c^2}{h'}\right)-1.
\label{O(N)contribution}
\ee
We substitute (\ref{dencoefftemp}), (\ref{O(N)contribution}) 
and (\ref{O(1)contribution}) into (\ref{denominator}), 
to obtain a simple expression of the denominator 
in terms of $R$:
\be
\int_{(x,y)\in\cS} dxdy \,e^{-\Veff (x,y)}
=(2\pi)^{\frac{3}{2}}\sqrt{(N-1)h'} \,
\exp\left(2(N-1)R+R+h'\frac{\der R}{\der h'}\right). 
\label{denominator_final} 
\ee
The first term in the exponential $2(N-1)R$ represents the leading part (\ref{denfinal}), 
and the remaining gives the next-to-leading contributions. 
A similar expression is obtained for the denominator 
in the definition of the chemical potential of an instanton 
in the standard $c<1$ noncritical string theory 
\cite{Ishibashi:2005zf}.  

Finally, we have a comment on the fact that 
${\cal O}(N^0)$ part in the exponent in the denominator 
(\ref{denominator_final}) can be written in terms 
of a cylinder amplitude. 
Noting that 
\be
h'\frac{\der}{\der h'}\vev{\cO}'=\prefII\vev{\cO\, \tr V(X')}'_{\rm conn},
\ee
where the subscript ``conn'' represents taking the connected part of the correlator, we find that 
\bea
h'\frac{\der R}{\der h'}
&=&\frac{1}{h'}\vev{\tr\log (\beta-X')\tr V(X')}'_c
-\frac{1}{2h'}\vev{\tr\log (c+1-\beta-X')\tr V(X')}'_c
+\frac{1}{2h'}V(\beta)\nono \\
&&+\left(\vev{\frac{1}{N-1}\tr\frac{1}{\beta-X'}}'_d
        +\frac{1}{2}\vev{\frac{1}{N-1}\tr\frac{1}{c+1-\beta-X'}}'_d
        -\frac{1}{2h'}V'(\beta)\right)h'\frac{\der\beta}{\der h'},
        \nono \\
\eea
where the last term vanishes due to (\ref{SPE}). 
Thus 
\bea
R+h'\frac{\der R}{\der h'}
&=&\vev{\frac{1}{N-1}\tr\log (\beta-X')}'_d
-\frac{1}{2}\vev{\frac{1}{N-1}\tr\log (c+1-\beta-X')}'_d
\nono \\
&&+\frac{1}{h'}\vev{\tr\log (\beta-X')\tr V(X')}'_c
-\frac{1}{2h'}\vev{\tr\log (c+1-\beta-X')\tr V(X')}'_c
\nono \\
&=& \int_{\Lambda}^{\beta}dz
    \left(R_{X'}(z)+\oint\frac{dz'}{2\pi i}\frac{1}{h'}V(z')
   \vev{\tr\frac{1}{z-X'}\tr\frac{1}{z'-X'}}'_c\right)
   \nono \\
&&-\frac{1}{2}\int_{\Lambda'}^{c+1-\beta}dz
    \left(R_{X'}(z)+\oint\frac{dz'}{2\pi i}\frac{1}{h'}V(z')
   \vev{\tr\frac{1}{z-X'}\tr\frac{1}{z'-X'}}'_c\right) \nono \\
 & & +\log\Lambda -\frac{1}{2}\log\Lambda'.
\label{denominator coeff}     
\eea
Namely, once we calculate the cylinder amplitude 
$\vev{\tr\frac{1}{z-X'}\tr\frac{1}{z'-X'}}'_c$, 
we should determine the ${\cal O}(N^0)$ coefficient in the denominator. 
In fact, the explicit form of this cylinder amplitude 
is given in \cite{Eynard:1995nv} for the $O(n)$ model with arbitrary $n$. 
(For the case $|n|<2$, see section 3.4 in the first paper of \cite{Eynard:1995nv}.)  
Using that expression in our case $n=1$, 
we find that the ${\cal O}(N^0)$ part in the exponent takes 
rather a simple form 
\be
R+h'\frac{\der R}{\der h'}
= \int_{-\frac{\Lambda}{h'}}^{z_0}dz
    \frac{1}{\sqrt{3}}G(z)+\log\Lambda
  -\frac{1}{2}\int_{-\frac{\Lambda'}{h'}}^{-z_0}dz
    \frac{1}{\sqrt{3}}G(z)
   -\frac{1}{2}\log\Lambda',
   \label{O(1) by G}
\ee
where $G(z)$ is a function introduced in \cite{Eynard:1995nv}, 
which is universal in the sense that it is uniquely determined 
only by the homogeneous saddle point equation 
(\ref{homogeneousSPE}) and by specifying its behavior near the edge of the cut 
and at the infinity. 
For details in the case $|n|<2$, see sections 3.2 and 3.3 in 
the first paper of \cite{Eynard:1995nv}. 
It is quite interesting that, even in the case of $c=0$ string 
theory defined by the one-matrix model, the denominator 
in the chemical potential of an instanton also has the next-to-leading term 
given by $G(z)$ for the $O(0)$ model 
as 
\be
R+h'\frac{\der R}{\der h'}
=\int_{\Lambda}^{\beta} dz
 \frac{1}{\sqrt{2}}G(z)+\log\Lambda
=\log\frac{\beta-\alpha}{4}. 
\label{O(0)}
\ee
It would be intriguing to examine whether this property holds 
for other $O(n)$ models. In our case (the $O(1)$ model), however 
complex form of $G(z)$ prevents us from performing 
the $z$-integration explicitly in (\ref{O(1) by G}) and 
we have not yet succeeded in obtaining a concrete value 
of the coefficient in the denominator. 
It would be natural to expect that it is somehow related 
to the length of the cut as in the case of the $O(0)$ model 
(\ref{O(0)}).

\end{document}